\documentclass{emulateapj}

\usepackage{subfigure,epsfig,url,bm,amsmath,cases,color,soul}

\begin{document}

\title{OSSOS XV: Probing the Distant Solar System with Observed Scattering TNOs}

\author{Nathan A. Kaib\altaffilmark{1}, Rosemary Pike\altaffilmark{2}, Samantha Lawler\altaffilmark{3}, Maya Kovalik\altaffilmark{4}, Christopher Brown\altaffilmark{1}, Mike Alexandersen\altaffilmark{2}, Michele T. Bannister\altaffilmark{5},  Brett J. Gladman\altaffilmark{6} \& Jean-Marc Petit\altaffilmark{7}}

\altaffiltext{1}{HL Dodge Department of Physics \& Astronomy, University of Oklahoma, Norman, OK 73019, USA}
\altaffiltext{2}{Institute of Astronomy and Astrophysics, Academia Sinica; 11F of AS/NTU Astronomy-Mathematics Building, No. 1 Roosevelt Rd., Sec. 4, Taipei 10617, Taiwan}
\altaffiltext{3}{Herzberg Astronomy and Astrophysics Research Centre, National Research Council of Canada, 5071 West Saanich Rd, Victoria, British Columbia V9E 2E7, Canada}
\altaffiltext{4}{Computer Science, Engineering, and Physics Department, University of Mary Hardin-Baylor, Belton, TX 76513, USA}
\altaffiltext{5}{Astrophysics Research Centre, Queen's University Belfast, Belfast BT7 1NN, United Kingdom}
\altaffiltext{6}{Department of Physics and Astronomy, University of British Columbia, Vancouver, BC V6T 1Z1, Canada}
\altaffiltext{7}{Institut UTINAM UMR6213, CNRS, Univ. Bourgogne Franche-ComtŽ, OSU Theta F-25000 Besan\c{c}on, France}

\begin{abstract}

Most known trans-Neptunian objects (TNOs) gravitationally scattering off the giant planets have orbital inclinations consistent with an origin from the classical Kuiper belt, but a small fraction of these ``scattering TNOs'' have inclinations that are far too large ($i>45^{\circ}$)  for this origin. These scattering outliers have previously been proposed to be interlopers from the Oort cloud or evidence of an undiscovered planet. Here we test these hypotheses using N-body simulations and the 69 centaurs and scattering TNOs detected in the Outer Solar Systems Origins Survey and its predecessors. We confirm that observed scattering objects cannot solely originate from the classical Kuiper belt, and we show that both the Oort cloud and a distant planet generate observable highly inclined scatterers. Although the number of highly inclined scatterers from the Oort Cloud is $\sim$3 times less than observed, Oort cloud enrichment from the Sun's galactic migration or birth cluster could resolve this. Meanwhile, a distant, low-eccentricity 5 M$_{\oplus}$ planet replicates the observed fraction of highly inclined scatterers, but the overall inclination distribution is more excited than observed. Furthermore, the distant planet generates a longitudinal asymmetry among detached TNOs that is less extreme than often presumed, and its direction reverses across the perihelion range spanned by known TNOs. More complete models that explore the dynamical origins of the planet are necessary to further study these features. With observational biases well-characterized, our work shows that the orbital distribution of detected scattering bodies is a powerful constraint on the unobserved distant solar system.

\end{abstract}

\section{Introduction}

While most trans-Neptunian objects (TNOs) reside on stable orbits that evolve very slowly, this is not true of scattering objects. Scattering objects are TNOs that exchange significant amounts of energy with the giant planets because they pass near one of the giant planets or chaotically evolve in a sea of overlapping resonances with Neptune \citep[e.g.][]{fern04, bailmal09}. \citet{glad08} define an object as scattering if its semimajor axis changes by at least $\pm1.5$ AU over the course of a 10-Myr integration with the giant planets. Because their dynamical lifetimes are much shorter than the age of the solar system \citep{dones96, tiscmal03, disbrun07}, scattering objects must be continually resupplied from some other region of the solar system, and the Kuiper belt is the most plausible and generally accepted source of these objects \citep{levdun97, dunlev97}. 

If the Kuiper belt is the source of all scattering objects, we should expect that the orbital inclinations of scattering objects will be comparable to the inclinations seen among the Kuiper belt because scattering interactions with the giant planets are not very effective at raising orbital inclinations \citep{brasser12}. While this is largely true, it is not always the case. For instance, \citet{glad09} reported the discovery of the first retrograde TNO, 2008 KV$_{42}$ (or Drac), which has an inclination of 103$^{\circ}$ and is actively scattering off of the giant planets ($q=21.1$ AU and $a=41.5$ AU). Similarly, \citet{chen16} announced the discovery of 2011 KT$_{19}$ (or Niku), another retrograde scattering TNO with an inclination of 110$^{\circ}$. These are two of the most extreme examples of highly inclined scattering TNOs, but numerous other scattering objects with $i>45^{\circ}$ can be found within the Minor Planet Center Database \citep[e.g.][]{brasser12, chen16}.

There is no known dynamical mechanism that can efficiently place objects from the classical Kuiper belt onto such highly inclined orbits \citep{brasser12, volkmal13}. However, processes outside the classical Kuiper belt may play a role. It has been suspected for some time that the Oort cloud may contribute to the observed scattering population \citep{emel05, kaib09, brasser12, gomes15}. The Oort cloud consists of a massive reservoir of distant ($a\gtrsim10^3$ AU) small bodies whose orbits are dynamically decoupled from the planets ($q\gtrsim45$ AU). In this scenario, galactic perturbations drive the perihelia of Oort cloud objects back into the planetary region, and energy kicks from planetary perturbations then draw the semimajor axes of these bodies to lower values \citep{bann17}. This dynamical process can occur at any orbital inclination, and because the Oort cloud's inclinations should be nearly isotropic, this will inevitably generate scattering objects with very high inclinations. 

Recently, however, a distant planet has also been proposed to orbit in the outer solar system with a semimajor axis of $\sim$500--1000 AU and an eccentricity of $\sim$0.2--0.7 \citep{batbrown16, bat19}. The existence of this planet has been suggested to explain the asymmetry in the orbital distribution of large semimajor axis ($a\gtrsim250$ AU) TNOs that are dynamically decoupled from the known planets ($q\gtrsim40$ AU) \citep{trujshep14, batbrown16}. Such a planet would enhance the production of scattering objects by driving the perihelia of distant TNOs into the planetary region (although the semimajor axes of the affected TNOs would generally be smaller than the semimajor axes of Oort cloud bodies) \citep{gomes15}. The mechanisms driving TNOs inward generally are secular resonances with the distant planet, and this dynamical process can also generate scattering objects with extreme inclinations \citep{batbrown16b, batmorb17, li18}.

While it is clear that the Oort cloud, as well as a distant planet, can generate a population of scattering bodies at high inclinations, it is not clear how abundant they should be relative to the population derived from the Kuiper belt. Moreover, it is difficult to use present observations to infer the intrinsic fraction of scattering objects residing at high inclinations. This is because the current sample of highly inclined scatterers is compiled from an amalgam of different TNO surveys. Typically, these surveys are concentrated along the ecliptic. Very roughly, this causes TNOs' discovery probabilities to scale with the inverse sine of their inclination (amplifying the dearth of known, highly inclined TNOs). However, precisely determining a TNO's discovery probability requires a thorough characterization of the surveys' observational biases, but for many surveys, these biases are undocumented or unknown. Although TNO orbits are listed in the Minor Planet Center Database, information on the pointing directions, sky coverage, magnitude limits, tracking fractions, etc. is often not provided for these discoveries. To this end, four TNO surveys over the past decade have been designed with the goal of carefully characterizing all survey biases: the Canada-France Ecliptic Plane Survey \citep[CFEPS; ][]{petit11}, its high-latitude extension \citep[HiLat; ][]{petit17}, the survey of \citet{alex16}, and the Outer Solar System Origins Survey \citep[OSSOS; ][]{bann18}. These surveys have discovered a total of over one thousand new TNOs with known discovery biases, allowing statistical testing of TNO orbital models \citep{law18c}.  Throughout the remainder of this paper we will refer to this group of four surveys as OSSOS+.

69 of the TNOs detected and tracked by OSSOS+ have been classified as scattering TNOs or centaurs (unstable objects with semimajor axes between the giant planets). This sample of objects has already been used to constrain the absolute magnitude ($H_r$) distribution of the low inclination scatterers, the bulk of which presumably come from the Kuiper belt \citep{shank13, shank16, law18}. In addition, the OSSOS+ catalog was already searched for signatures of a distant planet in the orbital distribution of scattering and detached TNOs, and no conclusive evidence was found \citep{law17}. However, the work's dynamical model used a distant planet orbit that was nearly coplanar with the known planets, and the search primarily focused on detected objects with high perihelion ($q>37$ AU) that were largely decoupled from the known planets. 

In the present work we use the OSSOS+ catalog of scattering TNOs to assess the different potential sources of high-inclination scattering objects. We first run three types of N-body simulations modeling the production of scattering objects from:  (1) the Kuiper belt, (2) the Kuiper belt and Oort cloud, and (3) the Kuiper belt and Oort cloud perturbed by an additional distant planet. By assuming an $H_r$ distribution for our N-body particles and simulating their detection within the OSSOS+ observing fields, we can assess how well each model replicates the actual OSSOS+ catalog of scattering objects and centaurs. Our work is organized into the following sections: In Section 2, we describe our numerical pipeline of N-body simulations and simulated survey detections. In Section 3, we present each dynamical model's simulated detections and compare them with the real OSSOS+ detections. Next, in Section 4, we discuss the implications of our work on the structure of the Oort cloud and the orbit and mass of any undiscovered distant planet. Finally, in Section 5 we summarize how the known high-inclination scattering TNOs offer a new probe of the very distant solar system.

\section{Dynamical Simulation Methods}

\citet{kaibshep16} ran four simulations of Kuiper belt formation to study the perihelion lifting and dynamical detachment that occurs within mean motion resonances (MMRs) of Neptune. These simulations, largely inspired by the works of \citet{nes15a, nes15b} and \citet{nes16}, migrated Neptune from 24 to 30 AU through a disk of 10$^6$ test particles. In all four simulations, when Neptune reached 28 AU, the migration was interrupted by a jump of 0.5 AU in Neptune's semimajor axis and 0.05 in eccentricity to mimic the effects of a gravitational scattering event with another giant planet \citep{nesmorb12}. In addition, the migration e-folding timescale was increased by a factor of $\sim$3 after the jump. Once Neptune reached its modern semimajor axis, the system was integrated until the $t=4.0$ Gyrs epoch with a timestep of 200 days. In two of the simulations, Neptune's migration was ``grainy,'' as thousands of small, instantaneous, random shifts of Neptune's semimajor axis of order $\delta a\sim10^{-3}$ AU were imposed throughout the migration to replicate the effects of scattering events with $\sim$2000 Pluto-mass objects.  

Of the four simulations run in \citet{kaibshep16}, we found that the ``Grainy Slow'' simulation best replicated the distribution of high-perihelion, high-inclination objects that are dynamically fossilized near Neptunian MMRs \citep{gom03}. This simulation, which we will refer to as GS16, employed a pre-jump migration e-folding timescale of 30 Myrs and a post-jump timescale of 100 Myrs (compared to our ``fast'' pre-jump and post-jump migration timescales of 10 and 30 Myrs, respectively). The parameters of this migration model largely agreed with the migration times and graininess levels favored in \citet{nes15a} and \citet{nes16} and also generated a high-perihelion, near resonant population consistent with the OSSOS dataset \citep{law18b}. \citet{nes16b} attained similar conclusions based on high-perihelion, near resonant TNOs. We refer readers to \citet{kaibshep16} for additional details about this simulation.

For the present work, we run three new simulations that are very similar to GS16. In the first new simulation (called ``OC''), the key difference is that this new simulation does not remove particles until they are 1 pc from the Sun. Because gravitational perturbations from passing field stars and the Galactic tide were omitted from \citet{kaibshep16}, GS16 removed particles at just 1000 AU. We now include both the Galactic tide and perturbations from passing field stars. Our prescription for the Galactic tidal force follows that of \citet{lev01} and includes a radial term derived from the Oort constants as well as a more powerful vertical term largely based on the local mean density of matter in the galactic disk, which we set to 0.1 M$_{\odot}$/pc$^3$ \citep{holmflynn00}. We assume a 60.2$^{\circ}$ inclination between the Galactic plane and the invariable plane of the solar system when calculating tidal terms. To limit the number of parameters varied in this work, this approach implicitly assumes that the Sun's galactic environment has remained fixed, but it is well-established that the Sun's stellar birth cluster and its migration within the Milky Way likely affects the Oort cloud's formation and enhances its population \citep[e.g.][]{bras06, kaib11}.

To model the effects of field star passages in our new OC simulation, we generate stellar-mass objects at randomly selected points 1 pc from the Sun. Based on the mass (spectral class) of each stellar object, we assign it a random velocity vector using the local solar neighborhood stellar dispersions and peculiar velocities given in \citet{gar01} and \citet{rick08}. Each star is then integrated in our simulation until its distance from the Sun exceeds 1 pc, at which point it is removed from the simulation. To determine the stellar passage rates of each spectral class of star, we assume the local density of stars in the solar neighborhood to be 0.034 M$_{\odot}$/pc$^3$, and we determine the spatial densities of each spectral class from the Present Day Mass Function given in \citet{reid02}. These densities combined with the spectral classes' velocity statistics allow us to set the encounter rate for each class. On average our simulations generate $\sim$18 stellar encounters within 1 pc of the Sun per Myr. This is comparable to Gaia-derived encounter rate of $20\pm2$ stellar passages within 1 pc per Myr \citep{bail18}.

Our next two new simulations, which we will call ``OC+P9a'' and ``OC+P9b'', are identical to OC except that the solar system also possesses an additional distant planet in orbit about the Sun. In OC+P9a, this planet is given a mass of 5 M$_{\oplus}$, a semimajor axis of 500 AU, an eccentricity of 0.25, and an inclination of 20$^{\circ}$. These choices are motivated by the recent analysis of \citet{bat19}. Meanwhile in OC+P9b, the distant planet is given a mass of 10 M$_{\oplus}$, a semimajor axis of 700 AU, an eccentricity of 0.6, and an inclination of 20$^{\circ}$. This choice of planet mass and orbit are based on the work of \citet{khain18}. Although attempts have been made to constrain the planet's other orbital elements \citep[e.g.][]{brownbat16}, these are not definitively known, and we randomly draw our planet's initial longitude of ascending node, argument of perihelion, and mean anomaly from uniform distributions. 

In our work here, we are interested in comparing the population of scattering objects generated in our simulations against those present in the OSSOS+ catalog. Following \citet{glad08} and \citet{shank13}, we define simulated objects as scattering if their semimajor axes change by more than $\pm$1.5 AU over the course of 10 Myrs of forward integration with the known giant planets. (This sample naturally includes the vast majority of objects that would also be classified as centaurs, which is why centaurs are included in our OSSOS+ observational sample.) For our GS16 and OC simulations, classifying scattering objects is simple, as we can just compare two simulation outputs that are separated by 10 Myrs. It is less straightforward for our OC+P9 simulations, however, since objects can scatter off of the known planets as well as our additional hypothetical planet or be moved into more strongly scattering orbits by the hypothetical planet's perihelion perturbations. When classifying real objects as scattering, only the gravitational effects of the known giant planets are considered. Therefore, to classify scattering objects in our OC+P9 simulations in a given time output, we take that time output, remove the distant planet, and integrate the augmented system for a separate 10-Myr interval while monitoring changes in test particle semimajor axes. The objects deemed to be scattering through this process comprise the scattering objects from our OC+P9 simulations.

\subsection{Survey Simulator}
\begin{figure*}[ht]
\centering
\includegraphics[scale=0.39]{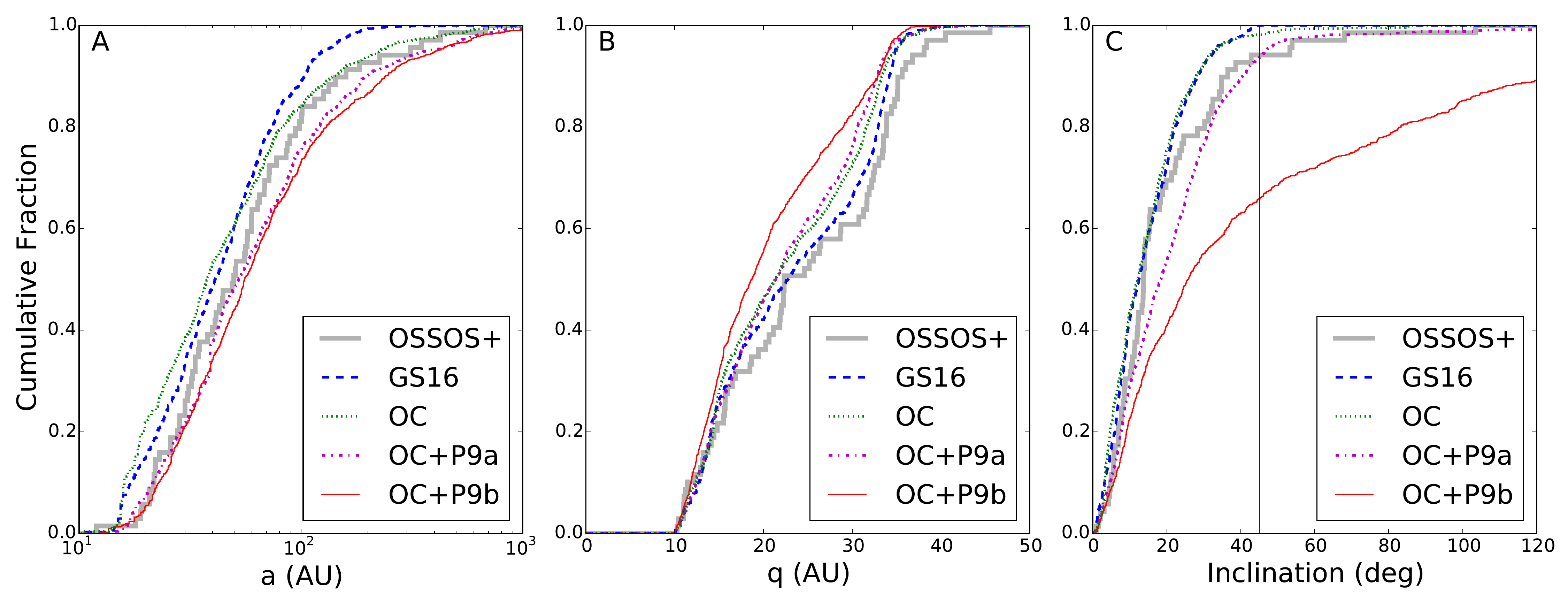}
\caption{Cumulative distributions of three orbital parameters for simulated scattering object detections using the GS16 ({\it blue dashed}), OC ({\it green dotted}), OC+P9a ({\it magenta dash-dot}) and OC+P9b ({\it red solid}) models, and the real OSSOS+ TNO detections ({\it grey solid}). Panel A shows semimajor axis, Panel B shows perihelion, and Panel C shows orbital inclination. The black vertical line marks an inclination of 45$^{\circ}$.}
\label{fig:modelcompare}
\end{figure*}

In order to directly compare our simulations with the OSSOS+ catalog we must first bias our dynamical simulation data to account for the simulated objects' differing discovery probabilities within these surveys. To do this, we employ the Survey Simulator constructed by the OSSOS and CFEPS teams \citep{law18c}. For a given simulated object, this simulator predicts whether it would be discovered by any one of the OSSOS+ affiliated well-characterized surveys. For each scattering object from our dynamical simulations, we first assign it an absolute magnitude ($H_r$) by randomly drawing from the best-fit ``divot'' $H_r$ distribution from \citet{law18}. The preferred distribution, derived from previous analysis of centaurs and scattering objects in the OSSOS+ catalog, is a disjointed differential power law ($\frac{dN}{dH_r} \propto 10^{\alpha}$) with a bright end ($H_r<8.3$) slope of $\alpha=0.9$ and a faint end slope of $\alpha=0.5$ \citep{shank13, shank16, law18}. ($H_r=8.3$ roughly corresponds to a physical diameter of 130 km assuming an albedo of 0.05.) In addition, the differential number discontinuously drops by a factor of 3.2 faintward of $H_r=8.3$ (the distribution's divot; see \citet{shank13} for details). In our work, we also try three alternative distributions to study how sensitive our results are to the selected size distribution. The first is the ``knee'' distribution of \citet{law18}, which has a bright-end power-law index of 0.9 that continuously (no divot) transitions to a faint-end power-law index of 0.4 at $H_r=7.7$. The second is the preferred double-power law of \citet{fras14}, which is identical to the \citet{law18} knee distribution except that the faint-end power-law is flatter with an index of 0.2. Our final alternative distribution is a single power-law that simply extends the bright 0.9 power-law to arbitrarily faint magnitudes. This last distribution has been ruled out with very high confidence \citep[e.g.][]{law18}, but we include it to explore our results' dependence on our assumed absolute magnitude distribution. 

Our combination of numerical simulations, $H_r$ distributions, and a survey simulator allows us to take a given model of the solar system's evolution and generate a synthetic catalog of scattering objects that would be found with the OSSOS+ ensemble of surveys. Comparisons between our simulated catalog and the real catalog of detected TNOs can then be used to statistically assess the success of a given solar system dynamical model at reproducing the real solar system.

\section{Results}

\subsection{Comparison of Simulated and Real Detections}

The OSSOS+ catalog contains 69 centaurs and scattering TNOs. 68 of these TNOs are listed in Table 3 of \citet{law18}. Since that publication, one additional object has been reclassified as scattering as its orbital measurements were refined (survey object name o5d144; $a=65.01$ AU, $q=34.43$ AU, and $i=13.81^{\circ}$). In Figure \ref{fig:modelcompare}, we compare the orbital distributions of 1000 simulated detections from each of our models with the 69 real scattering TNOs that are in the OSSOS+ catalog. One immediately obvious feature is that none of our models provide a perfect match to the inclination distribution of the real objects. While 4 of the 69 real scattering objects have inclinations above 45$^{\circ}$ (and a maximum of 103$^{\circ}$), the GS16 and OC models fail to replicate this high-$i$ fraction. The GS16 model yields no simulated detections above $\sim$45$^{\circ}$. Although the OC model does have a high-$i$ tail that extends all the way to retrograde inclinations, objects occupying this tail are still too rare and hard to detect, as only $1.8\pm0.4$\% of simulated OC detections have $i>45^{\circ}$, compared to $5.8\pm2.9$\% of real detections. 

In contrast, the OC+P9 models have the opposite problem; too many high-inclination scattering objects are generated. In the case of OC+P9b, this overproduction is egregious. $34.1\pm0.2$\% of the OC+P9b simulated detections have $i>45^{\circ}$ compared to just $\sim$6\% of real scattering TNO detections. Meanwhile, at first glance, the OC+P9a simulated detections appear to be a near-perfect match to the OSSOS+ dataset. $6.3\pm0.8$\% of simulated OC+P9a detections have $i>45^{\circ}$, effectively identical to the rate of $i>45^{\circ}$ objects observed in OSSOS+. However, the OC+P9a model also yields many simulated detections with moderate, but still significant inclinations. The median inclination of all OC+P9a simulated detections is 18.4$^{\circ}$, which is substantially higher than the 13.7$^{\circ}$ median of the actual OSSOS+ objects. \citet{nes17} have already discovered a similar issue with the population of Jupiter-Family Comets (JFCs). They find that a distant planet tends to generate JFCs with larger inclinations than those observed. Using a distant planet mass of 15 M$_{\oplus}$, the median inclination of JFCs produced in their model is $\sim$2$^{\circ}$ larger than the observed population, and they link this discrepancy to the inflated distribution of scattering objects' inclinations that the distant planet generates. Our deeper inspection of the scattering population confirms that this is indeed the case. Using a K-S test to compare observed JFC inclinations with those in their distant planet model, \citet{nes17} find a $p$-value of 0.008. Thus, they can reject the null hypothesis that observed JFCs and their simulated JFCs are drawn from the same distribution. When we perform the same K-S test comparing our simulated OC+P9a scattering detections and the real OSSOS+ scattering TNOs, we find a slightly smaller $p$-value of 0.006, rejecting the null hypothesis with similar confidence. (The same test comparing OC+P9b simulated detections and real OSSOS+ TNOs yields a $p$-value of $4\times10^{-6}$.)

\begin{figure} 
\centering
\includegraphics[scale=0.35]{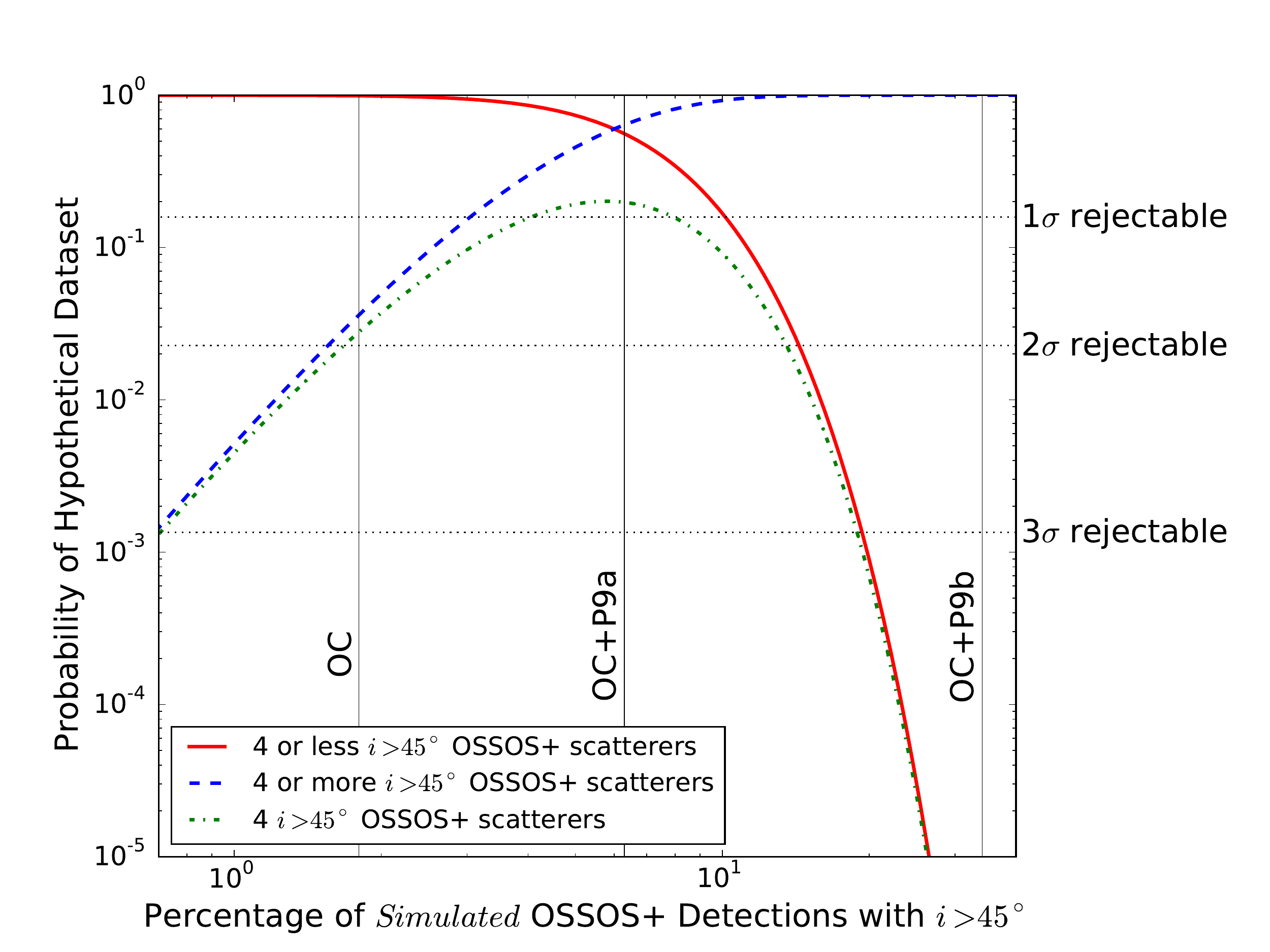}
\caption{The probability of the 69-object OSSOS+ scattering dataset occurring as a function of the ``intrinsic'' percentage of {\it simulated} detections that have $i>45^{\circ}$ for a hypothetical dynamical model and absolute magnitude combination. Green dash-dotted, red solid, and blue dashed lines respectively mark the probabilities of OSSOS+ finding 4 or less, 4 or more, and exactly 4 scattering objects with $i>45^{\circ}$ out of a total of 69 scattering detections. The ``intrinsic'' percentages of scattering detections with $i>45^{\circ}$ in our OC and OC+P9 models are marked with vertical lines and labeled accordingly. The right axis marks the confidence with which a model and $H_r$ distribution's intrinsic percentage can be rejected by the OSSOS+ catalog}
\label{fig:idealmodel}
\end{figure}

The failure of the GS16 model in Figure \ref{fig:modelcompare} shows that if we do not account for any distant perturbations on the solar system (Galactic tides and passing field stars in the OC model as well as a distant unseen planet in the OC+P9 models), we should essentially never expect scattering objects with inclinations over 45$^{\circ}$. Given this, there is a 0\% chance that the GS16 model can yield the actual OSSOS+ catalog where 4 of the 69 scattering objects have $i>45^{\circ}$. On the other hand, a model where 50\% of  simulated detections have $i>45^{\circ}$ would have an extremely tiny (but non-zero) probability of generating the OSSOS+ catalog since $\frac{4}{69}\ll0.5$. In Figure \ref{fig:idealmodel}, we plot the probability that a hypothetical dynamical model in combination with a given $H_r$ distribution combination will yield the number of $i>45^{\circ}$ scatterers detected in the OSSOS+ catalog. To do this, we assume that the model and $H_r$ distribution yield an ``intrinsic'' fraction of $i>45^{\circ}$ scattering detections that would be replicated if the OSSOS+ survey was carried out until a huge number of TNOs were detected. However, the actual OSSOS+ dataset is finite and thus subject to the uncertainties of small number statistics. Using the ``intrinsic'' $i>45^{\circ}$ detection fraction of a given model and $H_r$ distribution, we can then use the binomial distribution to estimate the chance that this model-distribution combination will generate the actual OSSOS+ results where 4 out of 69 detected scatterers are observed with $i>45^{\circ}$.

From Figure \ref{fig:idealmodel}, it is obvious that the GS16 model paired with the \citet{law18} divot $H_r$ distribution cannot explain the actual OSSOS+ dataset, since none of our simulated scattering detections have $i>45^{\circ}$. Meanwhile, it is also extremely unlikely for the OC+P9b model to generate the OSSOS+ scattering dataset. Figure \ref{fig:modelcompare} shows that 34\% of simulated scattering detections from this model should have $i>45^{\circ}$, and with this ``intrinsic'' fraction of highly inclined scatterers, Figure \ref{fig:idealmodel} indicates that the probability of OSSOS+ detecting 4 or fewer highly inclined scatterers out of 69 total objects is well below 1 in 10$^5$ assuming the OC+P9b underlying model. (It is actually $2.4 \times 10^{-8}$.) 

Our OC model is also unlikely to yield the actual OSSOS+ catalog of scatterers. The survey simulator predicts that 2\% of scattering detections will have $i>45^{\circ}$ when this model is paired with the \citet{law18} divot $H_r$ distribution. Figure \ref{fig:idealmodel} therefore indicates that there is only a 5.0\% chance that 4 or more OSSOS+ scatterers should have $i>45^{\circ}$, given this dynamical model and $H_r$ distribution. While this would still make the OSSOS+ catalog a $\sim$2-$\sigma$ outlier for our OC model, it is also $\sim$6 orders of magnitude more likely than the GS16 or OC+P9b models paired with the same $H_r$ distribution. Furthermore, while the OC+P9a model fares much better than the OC model in terms of the fraction of $i>45^{\circ}$ detections, the median inclination of OC simulated detections (12.7$^{\circ}$) is much closer to the median inclination of actual OSSOS+ detections (13.7$^{\circ}$). A K-S test that compares inclinations of simulated OC detections with real OSSOS+ scatterers returns a $p$-value of 0.19, indicating the null hypothesis cannot be rejected with as much confidence for the OC model compared to the OC+P9a model.

\begin{figure*}[ht]
\centering
\includegraphics[scale=0.47]{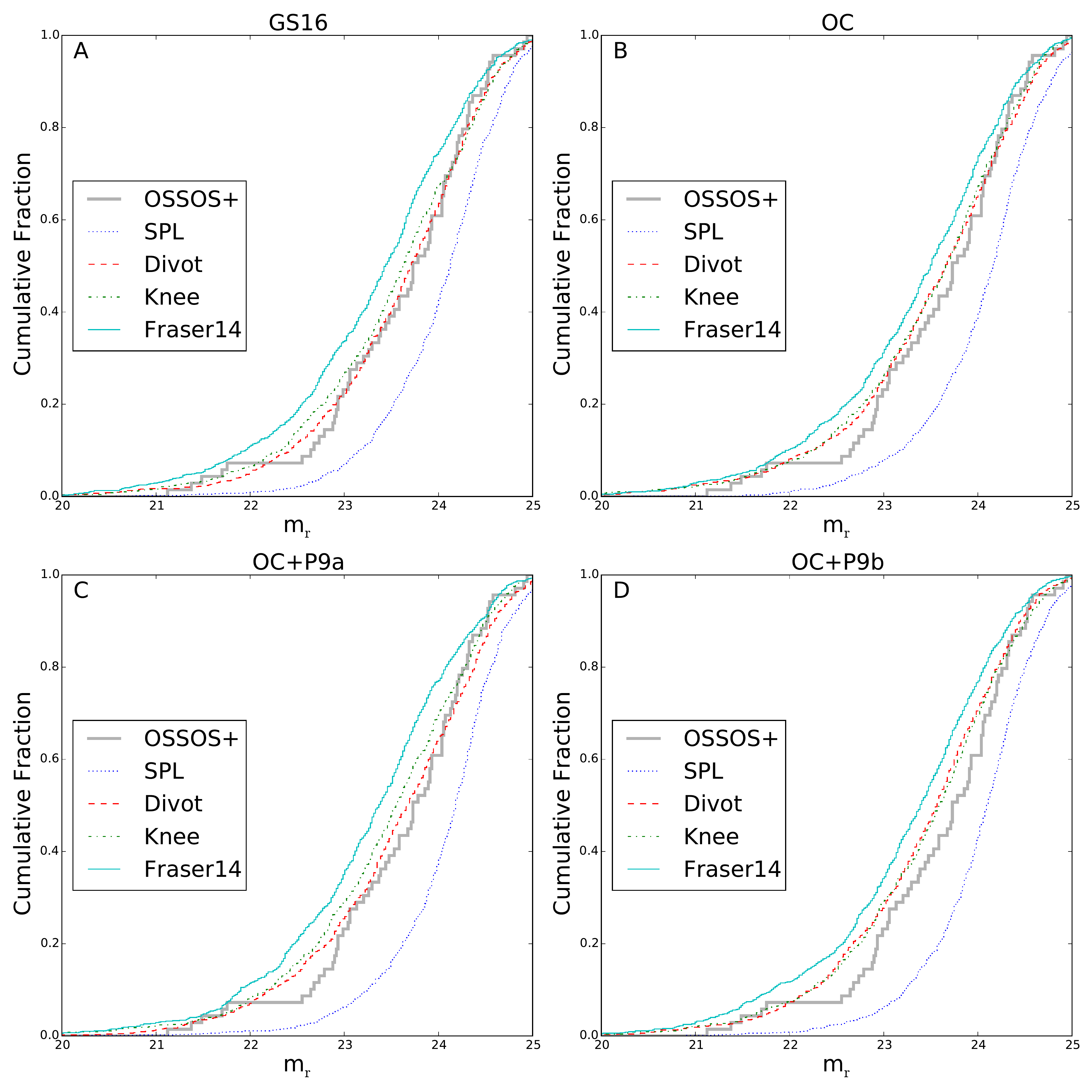}
\caption{{\bf A:}  The distribution of the apparent $r$-magnitudes of simulated scattering detections from the GS16 model. Sets of simulated detections are generated assuming the single power-law ({\it blue dotted}), \citet{law18} divot ({\it red dashed}), \citet{law18} knee ({\it green dash dotted}), and \citet{fras14} knee ({\it thin cyan solid}) $H_r$ distributions. The OSSOS+ scattering detections are shown by the thick gray solid line. {\bf B:} The distributions here are analogous to panel A, except the OC model is used. {\bf C:} The distributions shown here are analogous to panel A, except the OC+Pa model is used. {\bf D:} The distributions shown here are analogous to panel A, except the OC+Pb model is used.}
\label{fig:Comparemags}
\end{figure*}

It is important to emphasize that the results in Figure \ref{fig:modelcompare} employ the least rejectable ``divot'' absolute magnitude distribution from \citet{law18}. This $H_r$ distribution is derived through comparing only the low inclination members ($i<35^{\circ}$) of the scattering TNOs with a Kuiper belt formation model that is less complex than those explored here \citep{kaib11}. Because we are using new dynamical models of the Kuiper belt and also comparing them to all observed orbital inclinations, it is likely that the `best-fitting' absolute magnitude distribution will vary slightly from model to model and will also be somewhat different than \citet{law18}. However, the large majority of observed scattering objects are actually on low-inclination orbits. Furthermore, all dynamical models include an early large-scale scattering phase like the one employed in \citet{law18}, and because the dynamical lifetimes of scattering objects are short compared with the age of the solar system, most of the scattering objects present at the end of the simulation have recently left dynamically stable phase space, and thus have never ventured far from the Sun where they can experience strong perturbations. Thus, we expect the variations between the best-fitting absolute magnitude distributions of our new models and that of \citet{law18} to be minor. (Figure 3 of \citet{law18} gives an idea of the range of divots, knees, and break magnitudes that are statistically acceptable using a low inclination scattering model compared with OSSOS+.\footnote{It should be noted that the favored \citet{law18} absolute magnitude distribution is also consistent with constraints on the Plutino population's distribution \citep{alex16}.})

Nevertheless, it is important to better understand how sensitive our results are to our chosen absolute magnitude distribution. To do this, we repeat the survey simulations shown in Figure \ref{fig:modelcompare} three more times using three alternative $H_r$ distributions. The first alternative distribution is the preferred ``knee'' distribution from \citet{law18}. Our second distribution is the best-fit knee distribution found in \citet{fras14}. Finally, our last alternative distribution is a single power-law distribution (or SPL) with an index of 0.9 extending across all absolute magnitudes. A graphical representation of these different distributions can be found in Figure 2 of \citet{law18}.

In Figure \ref{fig:Comparemags}, we show the $r$-magnitude distributions of 1000 simulated detections from each of our models (GS16, OC, OC+P9a, and OC+P9b) when using our four different absolute magnitude distributions. Our various $H_r$ distributions primarily differ from one another at the faint end. As a result, the \citet{fras14} distribution ends up producing too many bright detections in each of our dynamical models, since it has a shallow power-law index at faint absolute magnitudes. In contrast, our single power-law distribution generates too many faint detections for every dynamical model, since the steep power-law index observed at brighter absolute magnitudes continues down to faint absolute magnitudes. Because our simulated detections bracket the apparent magnitudes of the actual detected objects, it is likely that the true ``best-fit'' absolute magnitude distribution of each dynamical model would yield detected orbits that fall within the range of those that our four different absolute magnitude distributions yield.

\begin{figure*} 
\centering
\includegraphics[scale=0.38]{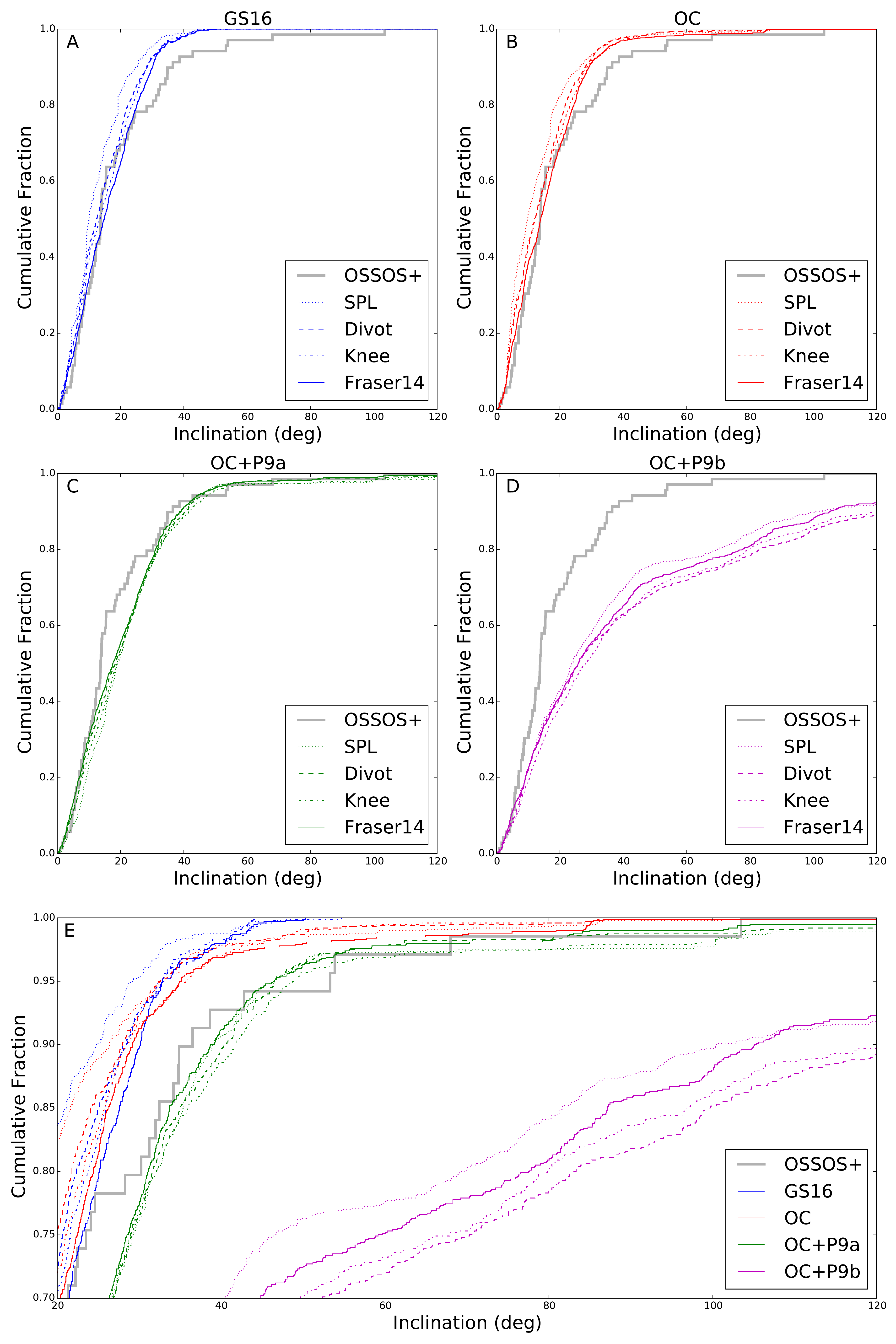}
\caption{{\bf A:}  The distribution of the orbital inclinations of simulated scattering detections from the GS16 model. Sets of simulated detections are generated assuming the single power-law ({\it dotted}), \citet{law18} divot ({\it dashed}), \citet{law18} knee ({\it dash dotted}), and \citet{fras14} knee ({\it thin solid}) $H_r$ distributions. The OSSOS+ scattering detections are shown by the thick gray line. {\bf B:} The distributions here are analogous to panel A, except the OC model is used. {\bf C:} The distributions shown here are analogous to panel A, except the OC+P9a model is used. {\bf D:} The distributions shown here are analogous to panel A, except the OC+P9b model is used. {\bf E:} A zoomed-in view of the high-inclination tail of the inclination distributions. GS16, OC, OC+P9a and OC+P9b distributions are marked with blue, red, green and magenta lines, respectively. The line styles have the same correspondences as in panels A--C.}
\label{fig:CompareIncs}
\end{figure*}

Now that we have established that our four possible absolute magnitude distributions likely bracket the best-fit absolute magnitude distribution in each dynamical model, we next study how these different distributions impact the detected inclinations that each dynamical model generates. In Figure \ref{fig:CompareIncs}, we show the distribution of simulated detected object inclinations in each of our models while assuming each of our four absolute magnitude distributions. Here we see that regardless of the $H_r$ distribution, observed inclinations above 45$^{\circ}$ are very unlikely within the GS16 model. Depending on the assumed $H_r$ distribution, only 0.0--0.5\% of simulated detections have inclinations over 45$^{\circ}$. However, the knee, divot, and \citet{fras14} distributions all nicely match the actual detected inclination distribution below $i<25^{\circ}$. On the other hand, the single power-law distribution generates an even greater number of low inclinations and is a poor match to the actual OSSOS+ catalog across a large range of inclinations. This is unsurprising since previous works have ruled out a single power-law distribution with a very high degree of confidence \citep{shank13, shank16, law18}.

Our four absolute magnitude distributions behave similarly in the OC model. The single power-law overproduces the number of detected bodies with low inclinations, while the knee, divot, and \citet{fras14} distributions yield better matches for $i<25^{\circ}$. Compared to the GS16 model, the OC model predicts a much higher fraction of detected bodies (1.5--2\%) will be on high inclination ($i>45^{\circ}$) orbits. While still 3--4 times rarer than the actual fraction (6\%) of detected high-inclination bodies, this is a much better match to the inclination data than the GS16 model. The differences between the simulated GS16 detections and simulated OC detections can be best seen in Figure \ref{fig:CompareIncs}E, which zooms in on the high-inclination tails of the distributions.

In our OC+P9a model, the assumed absolute magnitude distribution makes little difference to the inclinations of simulated detections. The median of the simulated detection inclinations varies between 17.6$^{\circ}$ and 19.0$^{\circ}$, or $\sim$4--5$^{\circ}$ greater than the median of actual detected inclinations, 13.7$^{\circ}$. In addition, the fraction of simulated detections above $i>45^{\circ}$ ranges from 6--8\%, compared to the actual fraction of 6\%. Thus, for any reasonable choice of absolute magnitude distribution, it appears to hold true that the OC+P9a model nicely matches the fraction of detections with $i>45^{\circ}$ but also yields too many scattering detections with moderate inclination. 

Similar to our OC+P9a model, the inclinations of OC+P9b simulated detections are fairly insensitive to the assumed absolute magnitude distribution.. The median of the simulated detection inclinations varies between 24.6$^{\circ}$ and 29.1$^{\circ}$, a factor of $\sim$2 greater than the median of actual detected inclinations, 13.7$^{\circ}$. In addition, a very large number of high-inclination detections are predicted for each assumed $H_r$ distribution. The fraction of simulated detections above $i>45^{\circ}$ ranges from 26--34\%, compared to the actual fraction of 6\%. Consequently, the inclination distribution of scattering objects in the OSSOS+ dataset rules out our OC+P9b model with a high degree of confidence ($\sim$99.999\%) for a wide range of absolute magnitude distributions.

\section{Discussion}

\subsection{Asymmetries of Distant Perihelion Directions}

One of the main motivations behind the hypothesis of a distant, undiscovered planet in the outer solar system is its ability to generate an anisotropic distribution in the perihelion directions of distant TNOs that are dynamically decoupled from the known planets \citep{trujshep14, batbrown16}. In the most recent analysis of distant known TNOs, \citet{bat19} find that the longitudes of perihelion ($\varpi$) of 11 of the 14 known TNOs with $q>30$ AU, $a>250$ AU, and $i<40^{\circ}$ are clustered within $\sim$90$^{\circ}$ of one another, while the remaining 3 constitute a second cluster directed $\sim$180$^{\circ}$ from the first cluster. However, the orbits of 5 of these 14 TNOs are also known to be rapidly diffusing under perturbations from the known planets. Removing these 5 unstable TNOs, the dominant cluster consists of 8 TNOs, and the less dominant is just represented with 1 TNO. It has been argued that this $\varpi$ asymmetry is a direct consequence of the dynamical influence of a distant, undetected planet on the distribution of distant, detached TNOs \citep{batbrown16}. 

With this in mind, it is important to measure the level of $\varpi$ anisotropy that is generated when the preferred Kuiper belt formation model of \citet{nes16} and \citet{kaibshep16} is executed in the presence of such a distant planet. We do so here using the raw results of the simulations and not considering observational bias, since OSSOS+ has only discovered 3 decoupled TNOs that would match the orbital cuts of \citet{bat19}. When measuring anisotropy of dynamically decoupled simulation particles, we crudely classify any object with $q>40$ AU as decoupled from the planets. (This is actually more stringent than the criterion of \citet{bat19} since we are considering extensive numerical simulations rather than the very limited known catalog of TNOs.) To quantify the level of anisotropy in our OC+P9 simulations, we just measure the number of decoupled objects with a longitude of perihelion ($\varpi$) within $\pm90^{\circ}$ of our additional planet (we consider these our ``aligned'' population) and compare it to the number of decoupled objects with a longitude of perihelion further than $\pm90^{\circ}$ from the planet (our ``anti-aligned'' population).

\begin{figure*}[ht]
\centering
\includegraphics[scale=0.38]{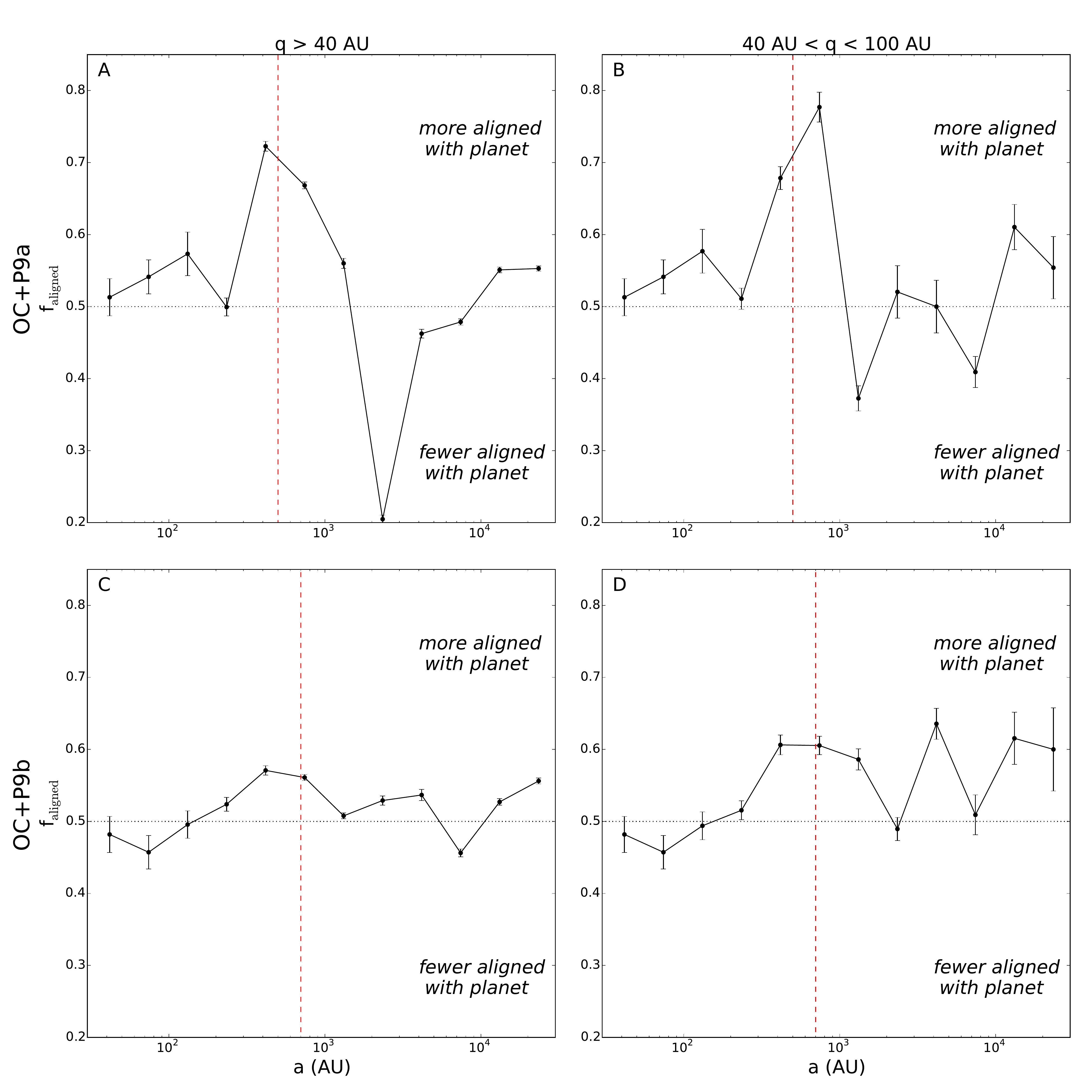}
\caption{{\bf A:} The fraction of orbits with $|\Delta\varpi|<90^{\circ}$ as a function of semimajor axis for all orbits with $q>40$ AU at the end of the OC+P9a simulation. $\Delta\varpi$ is the difference between an object's longitude of perihelion and the distant planet's longitude of perihelion. {\bf B:} To highlight the objects which are more likely to be detectable in a survey, here we show the fraction of orbits with $|\Delta\varpi|<90^{\circ}$ as a function of semimajor axis for all orbits with $40 < q < 100$ AU at the end of the OC+P9a simulation. {\bf C:} Analogous to Panel A, except the OC+P9b simulation is plotted. {\bf D:} Analogous to Panel B, except the OC+P9b simulation is plotted. In each panel, error bars mark 1-$\sigma$ Poisson uncertainties, the red dashed line marks the semimajor axis of the hypothetical distant planet, and the dotted line marks a symmetric split between aligned and anti-aligned longitudes of perihelion.}
\label{fig:lookatpomega}
\end{figure*}

In Figure \ref{fig:lookatpomega}A, we plot the fraction of $q>40$ AU objects that are in an aligned state as a function of their semimajor axis at the end of the OC+P9a model. For a truly isotropic population, this fraction should be 50\% according to the aligned criterion that we use. We see that for many semimajor axes the fraction does in fact stay within $\pm5$\% of this value. An exception to this occurs for decoupled objects with semimajor axes of 300--800 AU, near our distant planet's semimajor axis of 500 AU. For these objects, we see that there is a preference toward orbits that are aligned with the distant planet's longitude of perihelion. Surprisingly, the direction of this anisotropy is opposite to that found by previous studies of a distant planet's effects, which predict a strong preference for anti-aligned orbits \citep[e.g.][]{batbrown16, beck17, li18, had18}. Nevertheless, since we do not know the actual orbit of the distant planet (if it does in fact exist), at present we only consider the strength of the anisotropy. (We will address the anisotropy's direction in the next subsection.) 

As Figure \ref{fig:lookatpomega}A shows, there is also a range of semimajor axes that does possess a preference toward anti-aligned orbits. This occurs for semimajor axes between  $\sim$1000 and $\sim$4000 AU, where as many as 80\% of detached orbits are in an anti-aligned state. However, nearly all of the known TNOs that are thought to reflect the dynamical signature of a distant planet have semimajor axes below this range \citep{bat19}. 

One important caveat to the alignment fractions plotted in Figure \ref{fig:lookatpomega}A is that there is no upper limit on particles' perihelia. Because of this, many of these particles never make close approaches to the planetary system, which would dramatically lower their chance of detection. To ensure that the asymmetry displayed in Panel A would manifest itself in an observed sample, we restrict ourselves to decoupled objects with 40 AU $< q < 100$ AU in Figure \ref{fig:lookatpomega}B. Although detection probability is a very steep function of $q$, objects with perihelia approaching 100 AU have been detected \citep{brown04,trujshep14}, and we take this as the upper limit of a ``detectable'' set of orbits. In this new plot, we see that indeed the $\varpi$ asymmetry remains. For $40$ AU $< q<100$ AU and $300$ AU $< a<800$ AU orbits, the aligned fraction is 0.73, which corresponds to an aligned-to-anti-aligned population ratio of just under 2.7:1. This asymmetry is less dramatic than that inferred from TNO observations \citep[e.g.][]{trujshep14, batbrown16, bat19}. We also note that our aligned fraction in OC+P9a drops slightly to 0.67, or a 2:1 ratio, if we only consider orbits with $i<40^{\circ}$, as in \citet{bat19}.

In Figures  \ref{fig:lookatpomega}C--D, we look at the $\varpi$ alignment fraction at the end of model OC+P9b. Here we see that even though this model's distant planet is twice as massive (10 M$_{\oplus}$) as that used in OC+P9a (5 M$_{\oplus}$), the asymmetry in $\varpi$ is more mild. Among decoupled orbits ($q > 40$ AU) there is still a preference for aligned perihelion longitudes between $300$ AU $< a<800$ AU, but the aligned fraction is only $\sim$57\%. If we again restrict ourselves to particles with non-negligible chance of discovery ($40$ AU $< q<100$ AU), this aligned fraction rises modestly to $\sim$60\%, or a 1.5:1 ratio. Thus, the $\varpi$ asymmetry is weaker for very eccentric distant planet orbits, as suggested in \citet{bat19}. 

In the OC+P9 models, it appears that decoupled TNOs only have a $\sim$60--75\% probability of occupying an aligned state with the distant planet, which is not wildly higher than the 50\% probability that an isotropic model would give. It is therefore instructive to estimate how large of a sample size of distant, decoupled TNOs is necessary to confidently discern {\it our} OC+P9 models from models predicting an approximately isotropic distribution of decoupled orbits, such as Oort cloud formation within a birth cluster \citep{fern97, bras06}, Oort cloud formation nearer the Galactic center \citep{bras10, kaib11}, or mutual planetary embryo scattering \citep{gladchan06, siltre18}. Based on Figure \ref{fig:lookatpomega}, we assume that the ``intrinsic'' aligned $\varpi$ fractions are 0.73 and 0.6 for the OC+P9a and OC+P9b models, respectively. Then, assuming a hypothetical observed TNO sample size and a hypothetical observed aligned-to-anti-aligned ratio, one can use these intrinsic simulation fractions along with the binomial distribution to find the probability that this hypothetical observed aligned-to-anti-aligned ratio would arise within a randomly selected sample of our OC+P9 decoupled orbits. This probability can then also be computed for an isotropic distribution of decoupled orbits, which has an intrinsic aligned fraction of 50\%. If we then divide the OC+P9 probability by the sum of both probabilities, this then estimates the confidence with which an OC+P9 model is preferred over the isotropic model for the given observed sample size and observed alignment ratio.

\begin{figure} 
\centering
\includegraphics[scale=0.43]{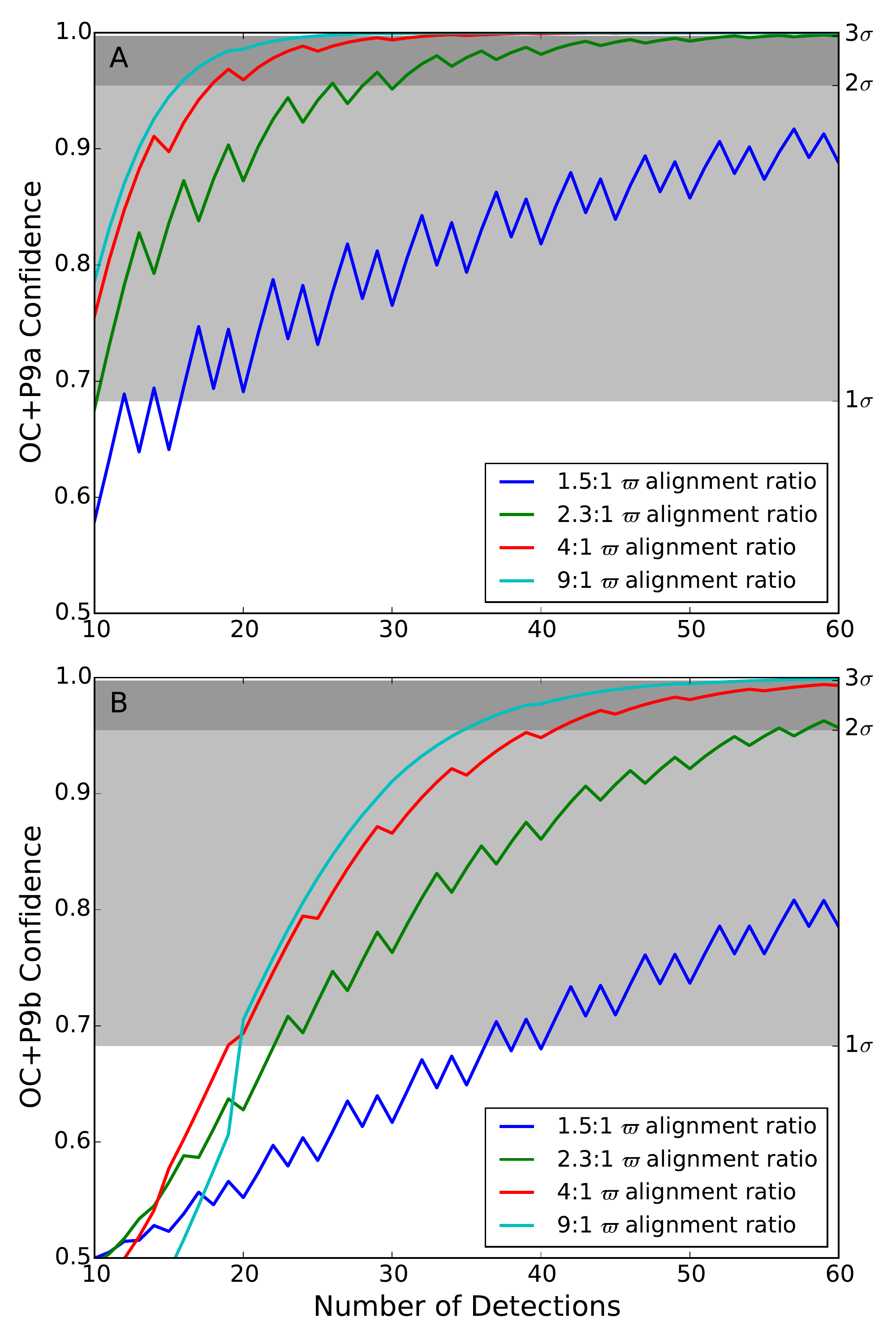}
\caption{{\bf A:} Confidence that our OC+P9a model is favored over an isotropic TNO perihelion distribution for a hypothetical observed $\varpi$ alignment ratio and a hypothetical observed TNO sample size. This confidence is plotted as a function of the total number of high-$q$ objects detected. The different color curves represent different observed aligned-to-anti-aligned ratios (see legend). From lightest to darkest, the shadings represent regimes where the OC+P9a model is preferred over an isotropic model at 0$\sigma$--1$\sigma$, 1$\sigma$--2$\sigma$, and 2$\sigma$--3$\sigma$ levels. {\bf B:} Analogous to Panel A, except the OC+P9b simulation is used.}
\label{fig:modconf}
\end{figure}

In Figures \ref{fig:modconf}A and \ref{fig:modconf}B we show the confidence that the OC+P9 models are preferred over an isotropic model as a function of object sample size for different observed alignment ratios. (The choppiness of the curves is due to the discreteness of the binomial probability distribution for small sample sizes.) As one can see in Panel A, distinguishing between the OC+P9a model and an isotropic model at a 1-$\sigma$ confidence can be done with a sample size of $\sim$10 or less objects. However, if one wishes to exceed 2-$\sigma$ confidence an observed dataset of 15--30 objects is needed, depending on whether the observed alignment ratio is 9:1, 4:1, or 2.3:1 (7:3). More sophisticated analyses of the orbital pole positions and eccentricity vectors of the $\sim$10 known decoupled high-$a$ TNOs suggest that the deviation from an isotropic model exceeds 3-$\sigma$ \citep{brownbat19}. However, Figure \ref{fig:modconf}A demonstrates that our OC+P9a model is not overwhelmingly favored over an isotropic model within the context of a simple binary grouping of $\varpi$ values and an observed sample size of $\sim$10 \citep{bat19}. This is simply because the $\varpi$ distribution of our OC+P9a model is not radically different from an isotropic one.

As Figure \ref{fig:modconf}B shows, confidently discerning the OC+P9b model requires an even larger observed sample of large semimajor axis, decoupled TNOs. To favor OC+P9b over an isotropic distribution at just 1-$\sigma$ confidence requires at least $\sim$20 detections for any of our tested observed alignment ratios (1.5:1 to 9:1). The current sample of TNOs used to study the asymmetry of decoupled objects varies somewhat between works, but is typically of order 10 objects \citep{trujshep14, batbrown16, sheptruj16, brown17}. Datasets where the OC+P9b model is preferred over an isotropic model with 2-$\sigma$ confidence require even larger sample sizes of 35--60 detected TNOs depending on the observed alignment ratio (or even larger still if the observed ratio is 1.5:1). 

The actual observed alignment ratio depends on what sample of decoupled TNOs is used, but recent works have cited alignment ratios as high as 9:1 \citep{brown17, shep18}. As shown in Figure \ref{fig:modconf}A, as long as the sample size is below $\sim$15 objects, the OC+P9a model is not favored over an isotropic model with a high degree of statistical confidence, and the OC+P9b is favored even more weakly. Of course, the reason that it is so difficult to discern our OC+P9 models from an isotropic distribution is that our OC+P9 models' alignment fractions are not extreme values very near 1. Models that generate a starker deviation from $\varpi$ isotropy would be favored more strongly by the current TNO sample. However, it appears to be difficult to generate an extreme $\varpi$ distribution if the detached TNO population is created through the distant planet's perturbations on a scattering population. \citet{nes17}, who performed similar simulations to ours, also report difficulty generating extremely asymmetric $\varpi$ distributions.

There are several possible explanations for the relatively modest $\varpi$ asymmetries in our OC+P9 models and the stronger asymmetries inferred from TNO samples. The first is that the observed alignment ratio is skewed because of a different dynamical mechanism at work that is not included in our simulations \citep[e.g.][]{mad16, sef19}. Another is that the observed alignment ratio is related to inherent observing biases. The well-characterized OSSOS+ survey results are in fact consistent with an isotropic distribution of perihelia \citep{law17}, and several of the observed TNOs used to motivate the existence of a distant planet were discovered in surveys without well-documented detection biases. If these unknown biases combine to enhance the detection probability of objects over a certain range of $\varpi$, this could artificially enhance the perceived anisotropy of the $\varpi$ distribution \citep{kav19}, although \citet{brownbat19} conclude that observing biases by themselves cannot fully explain the observed asymmetry. A final possible explanation is simply that our OC+P9 model is not a good analog for our solar system, and a better alternative evolutionary model is necessary to replicate the traditional Kuiper belt's observed structure \citep{bann18}, properly account for the number of high-inclination scatterers, and produce a stronger $\varpi$ anisotropy.

\subsubsection{Direction of $\varpi$ Asymmetry}

Our OC+P9 $\varpi$ distributions exhibit a glaring discrepancy with past works modeling the evolution of TNOs under the influence of a distant planet. While our and others' works find an anisotropic distribution of the longitudes of perihelion of detached ($q\gtrsim40$ AU) distant ($a\gtrsim300$ AU) particles, we find that there is an overabundance of particles whose longitudes are aligned within $\pm90^{\circ}$ of the distant planet's longitude. Meanwhile, most past works predict there should be a deficiency of such orbits and an overabundance of particles with longitudes that are anti-aligned with the planet's longitude.

We believe there are a couple of reasons for this discrepancy. The first is tied to our initial conditions. Our simulations begin with all of our particles on small semimajor axes ($a<30$ AU) and nearly circular orbits. To become dynamically detached from the known planets at semimajor axes of hundreds of AU, particles first have to undergo repeated interactions with the known planets to inflate their semimajor axes. Following this, particles with $a\sim10^{2-3}$ AU can subsequently acquire an extended perihelion distribution through interactions with the distant planet. In contrast, past works generally do not include these initial processes. Instead, their particles' initial orbital perihelia already extend to at least 50 AU, and the initial semimajor axes are in the hundreds of AU \citep[e.g.][]{batbrown16}. Often, these initially detached orbits are also isotropically distributed in argument of perihelion, longitude of ascending node, and, therefore, also longitude of perihelion. 

In our OC+P9 simulations, the same distant planet interactions that detach our particles' perihelia from the known planets simultaneously sculpt the newly detached particles' $\varpi$ distribution. This turns out to be a very important feature. To demonstrate this, we place $10^4$ particles on weakly detached (40 AU $<q<50$ AU) orbits with semimajor axes between 300 AU and 800 AU. These orbits' inclinations are distributed according to 

\begin{equation}
f(i) = \sin{i}\exp{-\frac{i}{2\sigma^2.}}
\end{equation}
where $\sigma$ is set to 15$^{\circ}$. Meanwhile, their arguments of perihelion and longitudes of ascending node are randomly drawn from an isotropic distribution. These particles are then integrated for 1 Myr under the gravitational influence of the Sun and known giant planets as well as the distant planet from our OC+P9a simulation ($m=5$ M$_{\oplus}$, $a=500$ AU, $e=0.25$, $i=20^{\circ}$, $\Omega=277^{\circ}$, and $\omega=323^{\circ}$). 

\begin{figure} 
\centering
\includegraphics[scale=0.43]{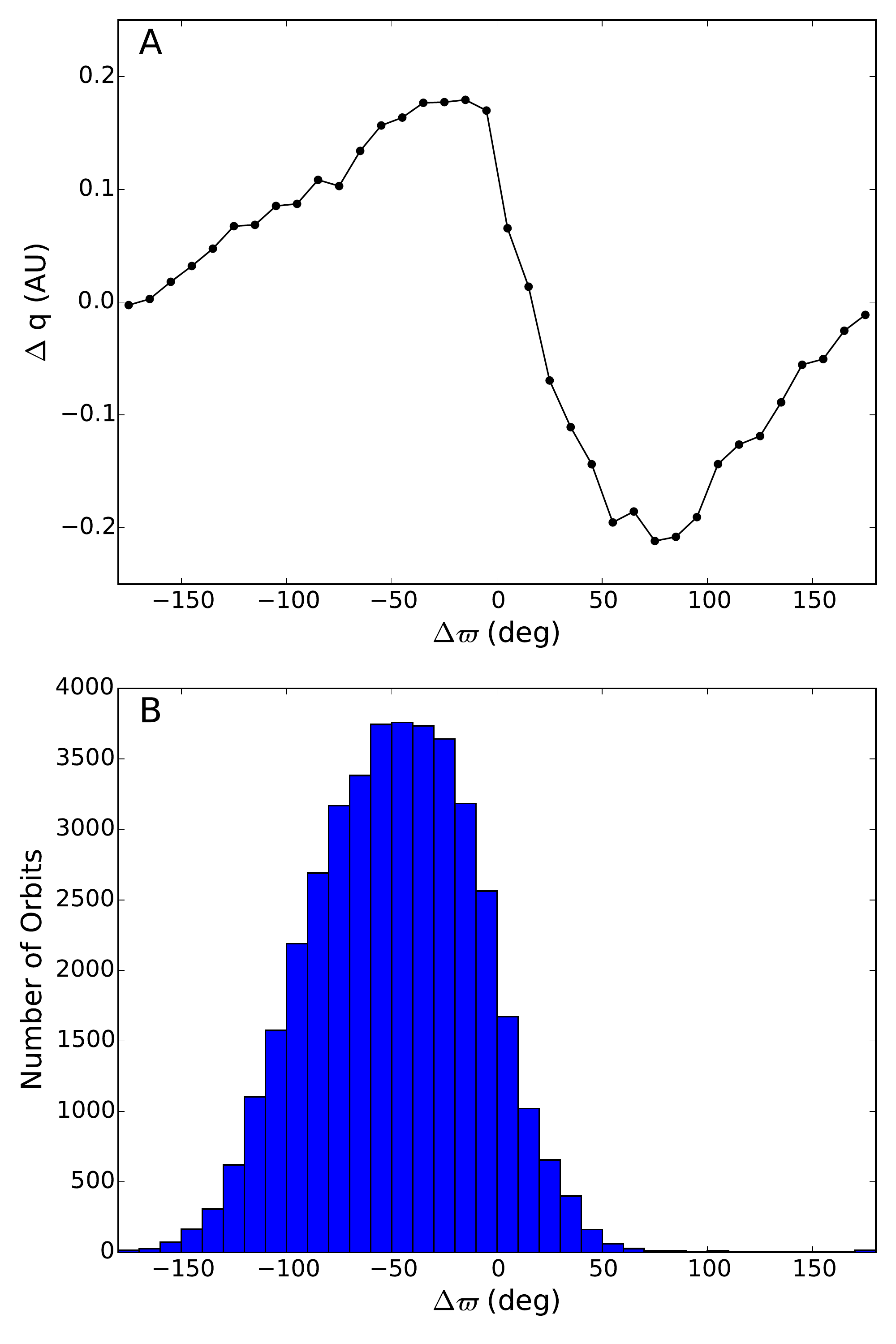}
\caption{{\bf A:} The median perihelion change over 1 Myr as a function of $\Delta\varpi$ for 10$^{4}$ particles with 40 AU $<q< 50$ AU and 300 AU $< a< 800$ AU. {\bf B:} Histogram of $\Delta\varpi$ values of $\sim$40,000 OC+P9a particles as they transitioned from $q<40$ AU to $q>40$ AU with semimajor axes between 300 and 800 AU.}
\label{fig:pomegaexplain}
\end{figure}

In Figure \ref{fig:pomegaexplain}A we plot the median change in perihelion of our particles over the course of our 1-Myr integration as a function of the difference between the particles' longitudes of perihelion and that of the distant planet, or $\Delta\varpi$. One sees that the typical perihelion change approaches 0 near $|\Delta\varpi|\simeq180^{\circ}$, indicating that our distant planet is ineffective at quickly detaching TNOs whose longitudes of perihelion are exactly anti-aligned with its own. Between $\Delta\varpi=-180^{\circ}$ and $\Delta\varpi\simeq0^{\circ}$, the median perihelion change steadily increases as $\Delta\varpi$ approaches 0$^{\circ}$. As a result, aligned orbits with $-90^{\circ}<\Delta\varpi<0^{\circ}$ are much more likely to become detached than anti-aligned orbits with $-180^{\circ}<\Delta\varpi<-90^{\circ}$. In addition, the sign of the perihelion forcing switches at $\Delta\varpi\simeq0^{\circ}$, and for $\Delta\varpi>0^{\circ}$, orbital perihelia tend to be driven closer to the known planets rather than detached. Consequently, the distant planet only efficiently detaches our particles' perihelia in the aligned quadrant of $-90^{\circ}<\Delta\varpi<0^{\circ}$. 

We can also verify that this is indeed what happens in our full OC+P9a simulation. Over the course of this 4-Gyr simulation, 40,024 particles with 300 AU $<a<800$ AU transition from $q<40$ AU to $q>40$ AU. In Figure \ref{fig:pomegaexplain}B we plot the distribution of $\Delta\varpi$ that these particles have when they make this transition. As can be seen, the distribution is dominated by $\Delta\varpi$ values between -90$^{\circ}$ and 0$^{\circ}$, and this range encompasses $\sim$74.6\% of the total distribution. It has been demonstrated that once an orbit is detached from the known planets, a distant planet can prevent its longitude of perihelion from fully circulating \citep{batbrown16, beck17, milllaugh17, shep19}. Therefore, since most particles first become detached from the known planets in a state that is aligned with the distant planet, it is no surprise that there is a surplus of aligned orbits at the end of our simulation. 

\begin{figure}
\centering
\includegraphics[scale=0.44]{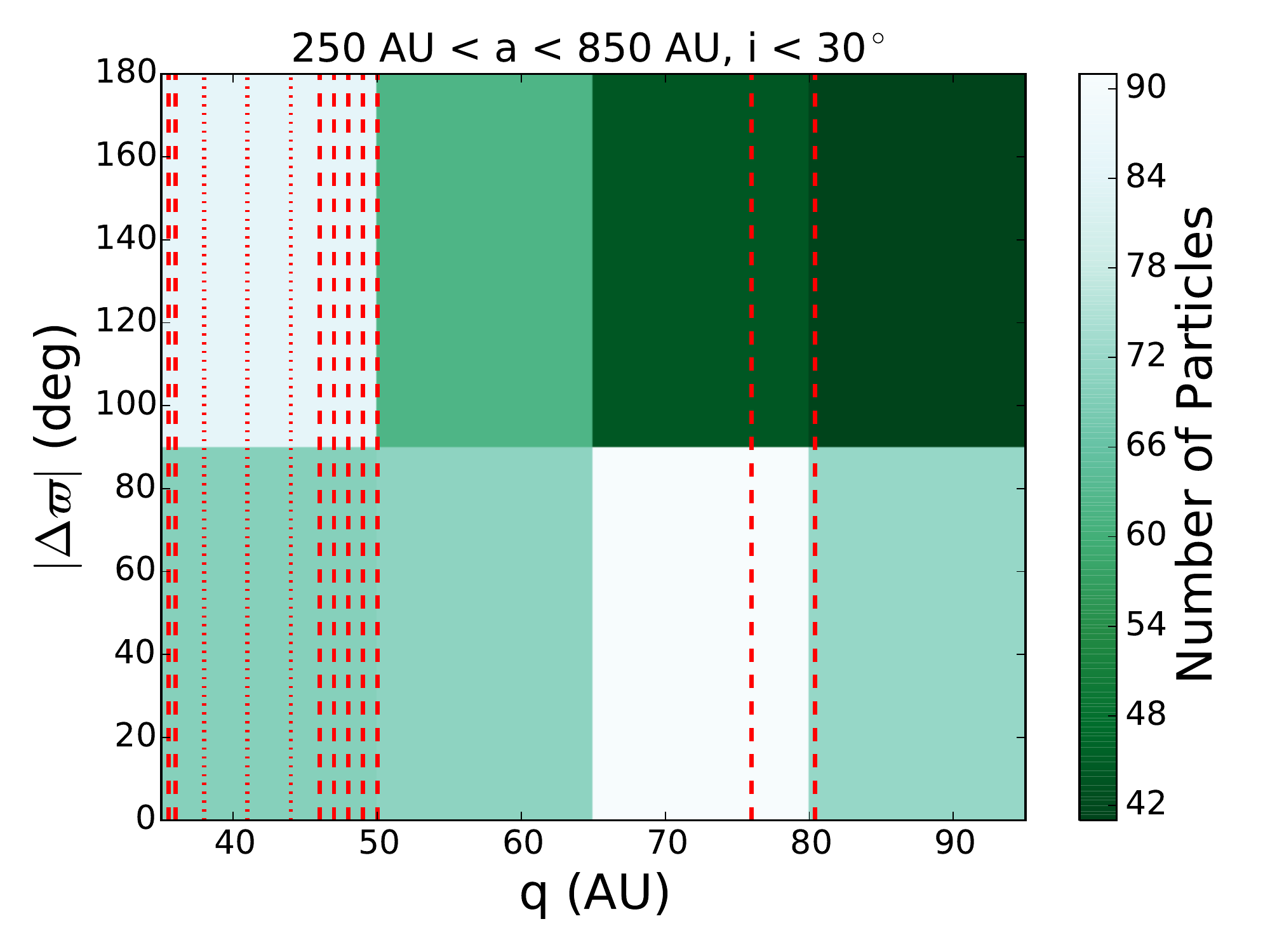}
\caption{Two-dimensional histogram of OC+P9a simulations particles' values of $|\Delta\varpi|$ and perihelion. Only particles with inclinations below 30$^{\circ}$ and semimajor axes between 250 AU and 850 AU are included. Dashed lines mark the known TNOs whose longitudes of perihelion are thought to be anti-aligned with a distant planet, while dotted lines mark known TNOs that are in a surmised aligned cluster. From left to right, vertical lines mark the perihelia of 2007 TG$_{422}$, 2013 RF$_{98}$, 2015 GT$_{50}$, 2015 KG$_{163}$, 2013 FT$_{28}$, 2015 RX$_{245}$, 2004 VN$_{112}$, 2014 SR$_{349}$, 2010 GB$_{174}$, 2013 SY$_{99}$, Sedna, and 2012 VP$_{113}$. }
\label{fig:qdelpomhist2d}
\end{figure}

Thus, our discrepancy with most past works is primarily explained by our simulations' absence of detached orbits at $t=0$. However, this does not explain the difference between our simulations' preference toward aligned orbits and the preference toward anti-aligned orbits reported in \citet{nes17}. The simulations performed in \citet{nes17} are very similar to ours; like us, they do not begin with detached orbits, yet they report the opposite $\varpi$ asymmetry that we do. However, their asymmetry is reported for orbits with 250 AU $<a<500$ AU, 35 AU $<q<50$ AU, and $i<20^{\circ}$. When we perform this same orbital cut, we do in fact see a preference toward anti-aligned values of $\varpi$. 66\% of such orbits are anti-aligned in our OC+P9a simulation, in agreement with \citet{nes17}. Upon further examination, we find that the preference in $\varpi$ is highly sensitive to perihelion values. In our OC+P9a simulation, this $\varpi$ preference is toward anti-alignment for $q\lesssim50$ AU and then rapidly switches to a significant alignment preference for $q\gtrsim50$ AU. 

In Figure \ref{fig:qdelpomhist2d}, we plot a 2-D histogram of particles' perihelia and $|\Delta\varpi|$ values in our OC+P9a simulation. Only particles with inclinations below 30$^{\circ}$ and semimajor axes between 250 AU and 850 AU are plotted. These orbital bounds encompass 12 of the 14 known TNOs that are used to examine $\varpi$ alignment in \citet{bat19}. (The two other TNOs, 2014 FE$_{72}$ and 2015 TG$_{387}$, have Oort cloud-like semimajor axes \citep{kaib11} and, according to Figure  \ref{fig:lookatpomega}, fall in a different orbital regime than the 250 AU $<a<850$ AU orbits.) The switch in $\varpi$ preference occurring near $q\simeq50$ AU can be readily seen in Figure \ref{fig:qdelpomhist2d}. This has major implications on the sample of known TNOs used to search for $\varpi$ clustering because two of most famous members, Sedna and 2012 VP$_{113}$ (and possibly 2013 SY$_{99}$ as well), sit in a perihelion range where the $\varpi$ preference should be opposite to the rest of the TNO sample used to search for $\varpi$ clustering. Within the context of our OC+P9a model, which assumes that a distant planet is responsible for sculpting the $\varpi$ distribution {\it and} dynamically detaching orbits from the known planets, Sedna and 2012 VP$_{113}$ should not be considered part of the same $\varpi$ clustering as the other TNOs discussed in \citet{bat19}. It is worth noting that these two objects both belong to the ``stable'' 9-object sample that most strongly exhibits $\varpi$ alignment, and their separate treatment would likely substantially alter the statistical significance of the observed $\varpi$ clustering. However, this separate treatment should only occur within models like OC+P9a and OC+P9b\footnote{In this subsection we have primarily focused on our OC+P9a simulation since it exhibits smaller inclination discrepancies with observed scattering TNOs, but we observe similar $\varpi$ trends in OC+P9b.} in which the distant planet is the mechanism that detaches TNO perihelia from the known planets, and other models that invoke another perihelion-detaching mechanism may not see such sensitive relationships between $\varpi$ clustering and perihelion.

Figure \ref{fig:pomegaexplain}A can explain the rapid switch at $q\simeq50$ AU from a preference toward anti-aligned $\varpi$ values to a preference toward aligned $\varpi$ values. Because the planet exerts little forcing on the perihelia of orbits with $|\Delta\varpi|\simeq180^{\circ}$, these orbits can remain in a weakly scattering  ($q\simeq35$ AU) or weakly detached ($q\sim40$--50 AU) state for a very long period of time. In contrast, orbits with $|\Delta\varpi|\simeq0^{\circ}$ experience stronger perihelion forcing, which drives them to a more detached state ($q\gtrsim50$ AU) or a more strongly scattering (and short-lived) state ($q\lesssim35$ AU). The shape and scale of the curve in Figure \ref{fig:pomegaexplain}A will undoubtedly vary for different masses and orbits of a distant planet. Nevertheless, if one assumes that the distant planet is the reason that Sedna-like objects have become dynamically detached from the known planets, then it is far from guaranteed that this planet will cause a preference toward anti-aligned longitudes of perihelion among all known detached, high-$a$ TNOs . 

\subsection{Nearly Coplanar Planet}

It is clear that our OC+P9 models generate scattering objects with overly excited inclinations and also do not generate extremely strong asymmetries in the distribution of the longitudes of perihelion of TNOs. One way to generate a stronger $\varpi$ asymmetry (without invoking a large, initial population of detached TNOs) may be to use a more massive distant planet. However, this will likely exacerbate the overproduction of high-inclination scatterers. 

Another potential solution that may not rely on tuning our particles' initial orbital distribution may be to employ a distant planet whose orbit is more coplanar with the known planets. While this paper is not an attempt to probe the full orbital and mass parameter space of distant planets to find the optimal combination, we can turn to previous work to see if a lower inclination distant planet can limit the production of high-inclination scatterers and generate a strong $\varpi$ asymmetry. To do this, we analyze a simulation from \citet{law17} that includes a 10 M$_{\oplus}$ distant planet on an orbit with $a=500$ AU, $e=0.5$, and $i=5^{\circ}$. In this simulation, the Kuiper belt is formed via the scattering of nearly circular, nearly coplanar test particles laid down between 4 and 40 AU, and no planetary migration is included. Thus, the model of Kuiper belt formation in \citet{law17} is inherently different from the simulations described in the earlier parts of our paper and would likely fail to replicate many observed features of the stable portions of the Kuiper belt \citep[e.g.][]{nes15a}. Nevertheless, this type of model does generate an acceptable low-inclination scattering population when a distant planet is excluded \citep{shank13}, and we can use this simulation to examine the effect of a nearly coplanar distant planet on the scattering population. 

\begin{figure} 
\centering
\includegraphics[scale=0.43]{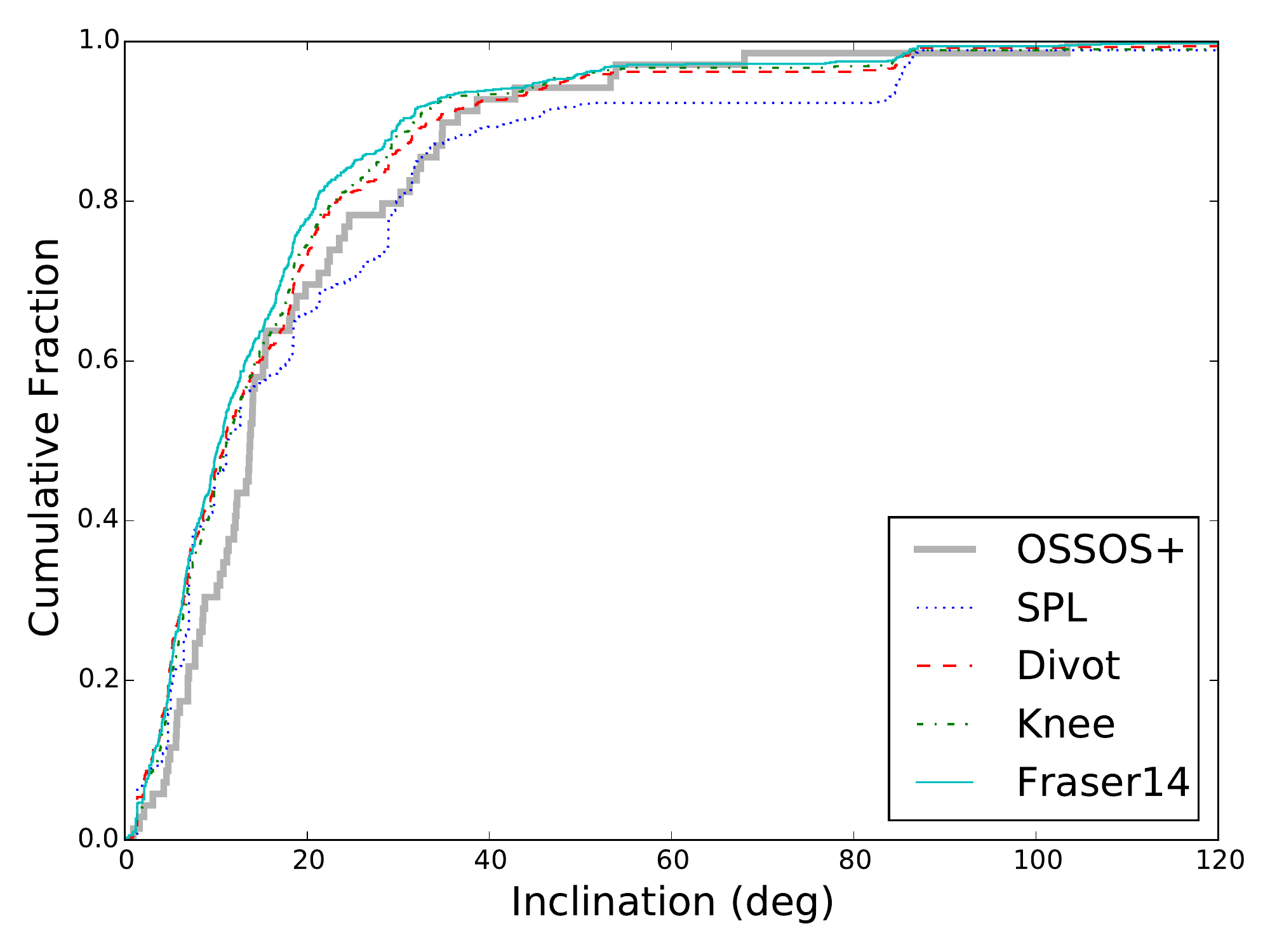}
\caption{The distribution of the orbital inclinations of simulated scattering detections from the \citet{law17} model that contained a low inclination ($i=5^{\circ}$) distant planet. Sets of simulated detections are generated assuming the single-power law ({\it blue dotted}), \citet{law18} divot ({\it red dashed}), \citet{law18} knee ({\it green dash dotted}), and \citet{fras14} knee ({\it thin cyan solid line}) $H_r$ distributions. The OSSOS+ scattering detections are shown by the thick gray line.}
\label{fig:LowincP9CompareIncs}
\end{figure}

In Figure \ref{fig:LowincP9CompareIncs}, we plot the distribution of detected OSSOS+ inclinations expected for scattering objects from the \citet{law17} model. To generate our underlying orbital distribution, we sample the simulation's scattering orbits every Myr for the final 100 Myrs of the simulation. Because this simulation contains 10 times fewer particles than the others explored in this paper, the distributions of detected inclinations are coarser. Nevertheless, Figure \ref{fig:LowincP9CompareIncs} shows that the \citet{law17} model provides a match to the actual OSSOS+ dataset that is superior to our OC+P9 models. Depending on the assumed absolute magnitude distribution, we expect 5--10\% of detected scattering objects to possess orbital inclinations over $45^{\circ}$. This compares very favorably with the 6\% of actual OSSOS+ scattering objects with $i>45^{\circ}$. The median detected inclination is between 10.5--11.3$^{\circ}$ depending on the assumed absolute magnitude distribution. This is less than the observed median of 13.7$^{\circ}$, but could easily be related to the fact that Neptunian migration is absent from this particular model \citep{nes15a}. 

While analysis of the full orbital elements of detached TNOs indicates that the nearly coplanar model is also rejected by the OSSOS+ dataset \citep{law17}, our analysis here is primarily focused on the inclinations of scattering TNOs. Figure \ref{fig:LowincP9CompareIncs} suggests that a distant, nearly coplanar, eccentric 10 M$_{\oplus}$ planet is not excluded by the OSSOS+ catalog of scattering objects, while a more highly inclined ($i\gtrsim20^{\circ}$) one is. Previous works have argued that the long-term torquing of a distant planet can explain the $\sim$6$^{\circ}$ inclination difference between the known planets' invariable plane and the Sun's equator \citep{bailey16}. However, this mechanism requires the hypothetical planet's inclination to be greater than $\sim$15$^{\circ}$ if its eccentricity is $\gtrsim$0.5 \citep{gom17}. While our nearly-coplanar planet results in an acceptable distribution of TNO inclinations, it is likely not the cause of the Sun's obliquity, and, indeed, other plausible explanations exist \citep[e.g.][]{lai11}.

Although a low-inclination distant planet may be consistent with the inclinations of detected scattering objects, the $\varpi$ asymmetry among decoupled ($q>40$ AU) objects with semimajor axis between 300--800 AU in our nearly coplanar model is still modest and virtually equivalent to the 60/40 split seen in our OC+P9b model. However, like our OC+P9 models, this one does not include a initially detached population of particles. Moreover, none of the models include a dynamical mechanism to place the distant planet on its orbit. Instead, the planet is assumed to have always been there, and a population of small bodies scattered from the known planets diffuses toward it. Its perturbations are not strong enough to produce an extreme orbital asymmetry within this scattered population, and the direction of the asymmetry that is generated swings wildly with small changes in perihelion. On the other hand, this distant planet's perturbations may be strong enough to carve an extreme $\varpi$ asymmetry within a population of decoupled objects not included in our models. If a dynamical mechanism such as a close stellar passage torqued the distant planet into the type of orbit explored here, it is quite possible that many small bodies were also torqued into detached orbits (and potentially with similar $\varpi$ alignments). It is still not clear, though, whether such a process will yield the level of asymmetry perceived among the modern catalog of TNOs, especially after this asymmetry becomes diluted by the nearly isotropic population generated from Neptunian migration during Kuiper belt formation. This scenario to generate greater $\varpi$ asymmetry is admittedly speculative and needs to be modeled in detail in future works.

Whether or not there is an additional population of decoupled small objects that are not included in our OC+P9 models, the distant planet will still perturb the solar system's known scattering disk and create the high-inclination scatterers we discuss in previous sections. Models including more decoupled objects to be influenced by the planet would likely only increase the number of high-inclination scatterers we report in this work. Thus, the overexcitation of scattering TNO inclinations should remain a feature of our OC+P9 models regardless of the assumed size of the initially detached TNO population.

\subsection{The Oort cloud as a Source of High-Inclination Scatterers}

We now return to further considerations of the Oort cloud as the main source of high-inclination scattering TNOs. As demonstrated in Figure \ref{fig:modelcompare}, detectable high-inclination scattering objects are produced in the OC simulation, but they are produced at a significantly lower rate (3--4 times) than what is observed. With this in mind, we now study the ultimate source region of these objects before they become detectable scattering objects in the final 500-Myr epoch of our simulation. To do this, we identify every scattering object in the final 500 Myrs that has $i > 45^{\circ}$ and $q>10$ AU and $a<1000$ AU. Then we study the dynamical history of these particles. 54\% of these bodies attained their large inclinations without ever actually spending any time in the Oort cloud (which we define as $q>45$ AU and $a>300$ AU; \cite{bras15}) prior to their scattering phase during the final 500 Myrs. Instead, most of these bodies attain a high inclination because they possessed a large ($\gtrsim10^3$ AU) semimajor axis during a previous epoch of scattering (sometime before the final 500 Myrs), allowing galactic perturbations to inflate their inclinations while their perihelia always remained near the known planets \citep[e.g.][]{lev06}. However, many of the objects with this type of dynamical history still do not significantly exceed 45$^{\circ}$ in the final 500 Myrs. Our sample of simulated objects with these dynamical histories has a median inclination of 52$^{\circ}$, whereas the four high-inclination OSSOS+ scattering objects all have inclinations above 53$^{\circ}$.

When we instead examine the dynamical histories of simulated scattering objects with $i\geq53^{\circ}$, we find that the Oort cloud becomes the dominant supplier of the scatterers. 59\% of these scattering objects previously resided in the Oort cloud before their scattering phase during the final 500 Myrs (while the remainder still attained their high inclination when they possessed a temporarily large heliocentric distance during an earlier scattering phase). Thus, the Oort cloud could be an important, and perhaps the dominant, source of the high-inclination scatterers detected in OSSOS+. 

\begin{figure} 
\centering
\includegraphics[scale=0.43]{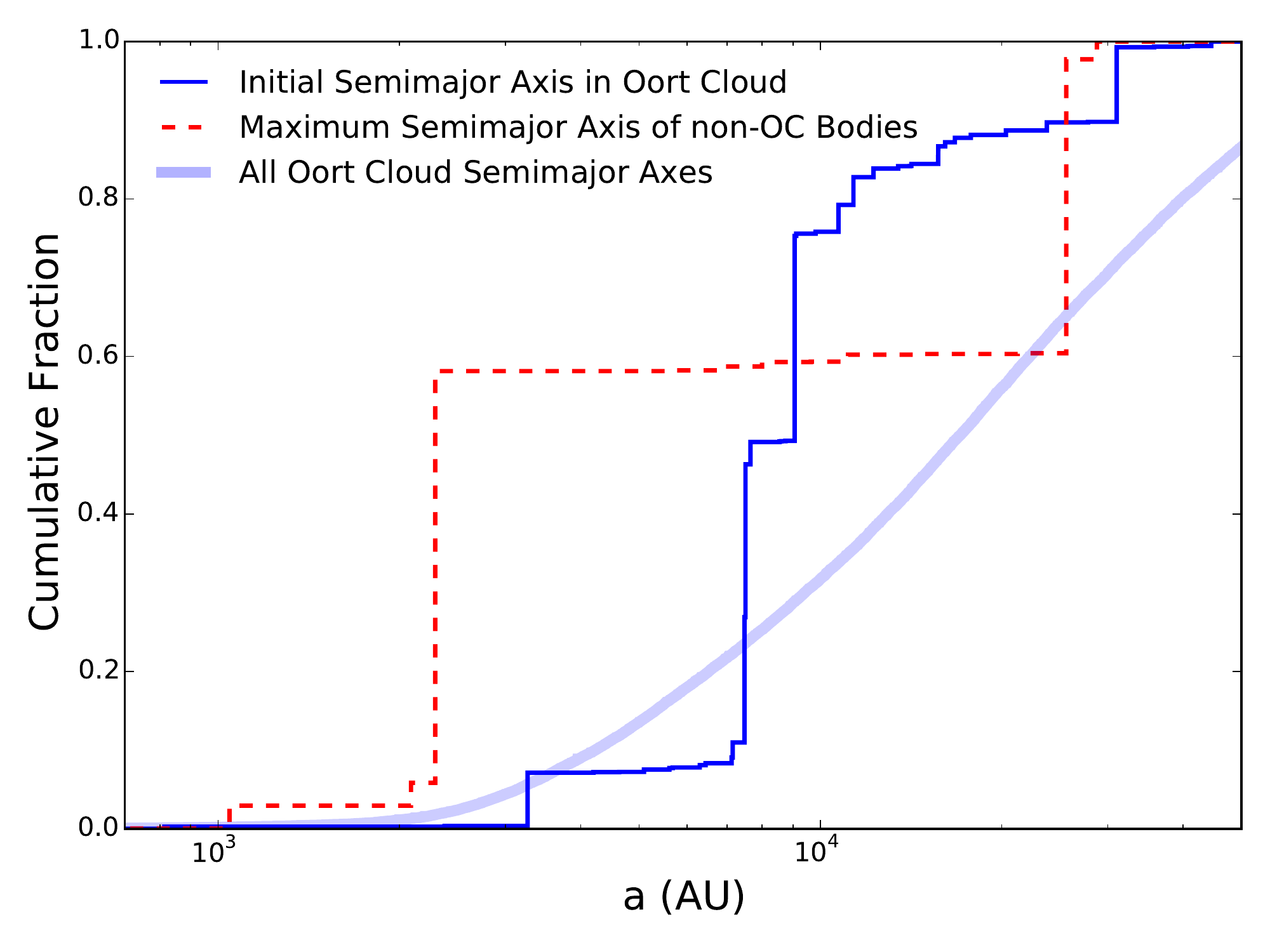}
\caption{Dynamical history information on scattering objects with $i\geq53^{\circ}$ from the final 500 Myrs of our OC simulation that are detected by the survey simulator. (The $i\geq53^{\circ}$ cut is used since all four high-inclination scatterers within OSSOS+ have $i\geq53^{\circ}$.) For simulated detected scattering objects that previously resided in the Oort cloud, we plot the cumulative distribution of the semimajor axes they had when they entered the Oort cloud in blue. For all other simulated detected scattering objects, we plot the cumulative distribution of the maximum semimajor axes they attain {\it before} they are recorded as scattering during the final 500 Myrs ({\it red dashed}). The thicker light blue line marks the {\it unbiased} distribution of all Oort cloud bodies in our OC simulation at $t=4$ Gyrs, where we define Oort cloud members as any bodies with $q>45$ AU and $a>300$ AU.}
\label{fig:highinchist}
\end{figure}

To better understand the production of detected high-inclination scattering objects, we run the OC model's distribution of high-inclination scattering orbits through the OSSOS+ survey simulator assuming the \citet{law18} divot absolute magnitude distribution. This is done until we compile $10^4$ detections. This sample of orbits now accounts for observing bias (whereas the dynamical histories discussed in the previous two paragraphs were all weighted equally). For the detections with $i\geq53^{\circ}$, we find that 56\% of objects resided in the Oort cloud in their past, and the other 44\% did not. In Figure \ref{fig:highinchist}, we plot the distribution of maximum semimajor axes that the non-Oort Cloud bodies attain prior to being recorded as scattering objects during the last 500 Myrs. As can be seen, all of these objects had attained semimajor axes beyond $\sim$2000 AU in a prior epoch. This allows galactic perturbations to torque their inclinations. Although these perturbations can also torque the perihelia out of the planetary region and into the Oort cloud \citep{oort50}, this did not occur for this subpopulation of bodies. One also notices that the distribution of maximum semimajor axes is not very smooth. This is because two objects (with maximum semimajor axes of $\sim$2000 and $\sim$26000 AU) spend time in scattering orbits that have very high detection probabilities, and their maximum semimajor axes consequently dominate the distribution.

Figure \ref{fig:highinchist} also provides information on the dynamical histories of detected high-inclination scattering objects that did previously reside in the Oort Cloud. For these bodies, we plot the distribution of semimajor axes they possessed when they first entered the Oort cloud ($q>45$ AU and $a>300$ AU). As can be seen, the inner regions of the Oort cloud dominate the production of detectable high-inclination scattering objects. These bodies first enter the Oort cloud with a median semimajor axis of 9000 AU. Meanwhile, if we look at the semimajor axes of all Oort cloud bodies at $t=4$ Gyrs, we find the median semimajor axis is nearly twice as large, or 17500 AU. This is unsurprising because in order to have a significant chance of detection, planetary perturbations must draw down the Oort cloud objects to small semimajor axes that enhance their detection probability \citep{kaib09}. The perihelia of Oort cloud objects with semimajor axes beyond $\sim$$2\times10^4$ AU can slide through the giant planets' region in a single orbital revolution, but the perihelia of smaller semimajor axis orbits cannot \citep{hills81, heitre86}. This results in a scattering phase for these smaller semimajor axis Oort cloud bodies while their perihelia are in planet-crossing configurations \citep{kaib09, kaibquinn09}, and this region of the Oort cloud therefore dominates the production of high-inclination scattering detections.

Thus, the production rate of high-inclination scattering objects is quite sensitive to the Oort cloud population between 1000 AU $\gtrsim a \gtrsim 2\times10^4$ AU. A boost in the number of objects residing here would also give a large boost to the number of detectable high-inclination scattering bodies. Such a scenario is plausible because the population size of this region is not well-constrained \citep{kaibquinn09} and is quite sensitive to the Sun's particular dynamical history. Simulations of the stellar dynamics of Milky Way-type galaxies show that stars often radially migrate away from the Galactic center on Gyr-timescales \citep{sellbin02, rosk08, loeb16}. This implies that the Sun once likely inhabited denser regions of the Galaxy, and these enhanced perturbations can enrich the subpopulation of the Oort cloud (1000 AU $< a < 2\times10^4$ AU) most likely to produce high-inclination scatterers. Previous work suggests that a strong outward radial migration history of the Sun can cause the size of this Oort cloud subpopulation to be 2--3 times larger than that predicted by models like our OC model that do not account for radial migration \citep{kaib11}. Such an enrichment would drive the OC model into a much closer agreement with the OSSOS+ catalog (see Figure \ref{fig:idealmodel}).

Of course, this region of the Oort cloud can be enriched in other ways as well. If the Sun formed as a member of an embedded star cluster, we would expect it to have spent its first $\sim$10 Myrs in a local environment whose perturbations were much more powerful than the current solar neighborhood \citep{lala03, fern97}. For clusters of moderate density, these perturbations will greatly enhance the population of the Oort cloud orbiting at semimajor axes of $\sim$10$^{3-4}$ \citep{bras06, kaibquinn08, bras12}. Again, this is the most productive region of the Oort cloud for high-inclination scatterers, and this population enhancement could likely explain the number of high-inclination scattering objects detected in OSSOS+.

\begin{figure} 
\centering
\includegraphics[scale=0.43]{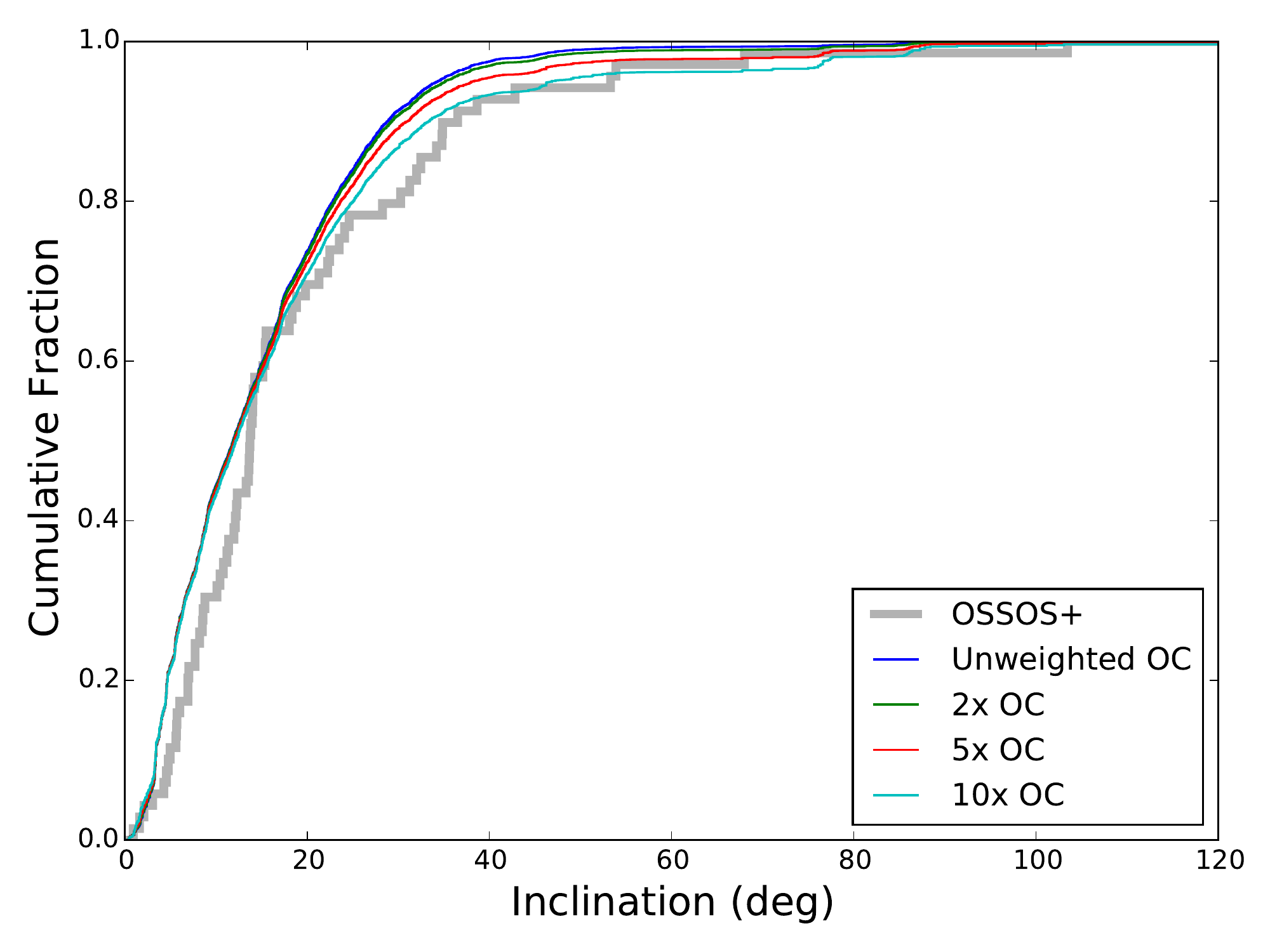}
\caption{The cumulative inclination distribution of scattering objects in the OSSOS+ dataset is shown with the thick gray line. Assuming the \citet{law18} divot $H_r$ distribution, the inclination distribution of simulated detections from the OC model is also shown with the other four lines. Scattering objects that come from the Oort cloud and originally entered the Oort cloud with semimajor axes below $2\times10^4$ AU are weighted by factors of 1, 2, 5, and 10 in the blue, green, red, and cyan distributions, respectively.}
\label{fig:IOCweights}
\end{figure}

In Figure \ref{fig:IOCweights} we demonstrate how the enrichment of the inner $\sim$$2\times10^4$ AU of the Oort cloud can provide a better match to the OSSOS+ scattering detections. Here we simply plot the cumulative inclination distribution of the simulated detections of the OC model when assuming the \citet{law18} divot $H_r$ distribution. For scattering objects that previously spent time in the Oort cloud and entered the Oort cloud with semimajor axes below $2\times10^4$ AU, we increase their weighting by factors of 2, 5, and 10 to approximate the expected inclination distribution when the inner Oort cloud population is enriched by these factors. As Figure \ref{fig:IOCweights} shows, it takes an order of magnitude population enhancement before the expected fraction of $i>45^{\circ}$ scattering detections matches the OSSOS+ dataset fraction (6\%). However, according to Figures \ref{fig:idealmodel} and \ref{fig:IOCweights}, the OSSOS+ dataset would be within 1$\sigma$ of the expected number of high-inclination scatterers for a factor of 5 population enhancement. Moreover, even for a factor of 2 inner Oort cloud population enhancement, the OSSOS+ dataset is well under a 2$\sigma$ outlier. Thus, the Oort cloud is a promising explanation for the high inclination scattering objects detected within OSSOS+.

\section{Conclusions}

The inclination distribution of centaurs and actively scattering TNOs discovered with OSSOS, CFEPS, the HiLat extension, and the \citet{alex16} survey contains a tail of very high inclination orbits, with 6\% of detected scattering orbits having $i>45^{\circ}$. No matter what absolute magnitude distribution we assume for the scattering population, this high-inclination tail is not reproduced within a Kuiper belt formation model that neglects the Oort cloud and only considers the known giant planets and their migratory histories \citep[e.g.][]{kaibshep16, nes16}. If the perturbations of passing field stars and the Milky Way tide are included in such models, the Oort cloud that consequently forms within these models will supply a steady population of high-inclination scattering orbits in the outer solar system. Within such models, we predict that $\sim$2\% of the scattering objects detected with the aforementioned surveys (OSSOS+) will be found with inclinations above 45$^{\circ}$. Although this fraction is a factor of $\sim$3 less than the observed fraction, the interior of the Oort cloud (1000 AU $< a < 2\times10^4$ AU) is a major supplier of these objects. It is well-known from previous studies of Oort cloud formation that the population size of this region of the Oort cloud can vary by a factor of several depending on the Sun's birth cluster and its radial migration history within the Milky Way \citep[e.g.][]{bras06, kaib11}. These effects are not included in the dynamical models presented here, and their inclusion would bring our dynamical models into closer agreement with the observed catalog of scattering objects.

We also examine the effects of a distant planet on the population of detected scattering objects by repeating our simulation two more times: once with a 5 M$_{\oplus}$ distant planet included on an orbit with $a=500$ AU, $e=0.25$, and $i=20^{\circ}$, and a second time with a 10 M$_{\oplus}$ distant planet included on an orbit with $a=700$ AU, $e=0.6$, and $i=20^{\circ}$. Of the two hypothetical planets, we find that the less massive, more circular planet provides a better match to observed scattering inclinations, which is consistent with other independent analyses \citep{bat19}. While this hypothetical planet produces a suitable number of very high ($i>45^{\circ}$) scattering objects, the overall median inclination of simulated scattering detections is 4--5$^{\circ}$ higher than the actual observed value. Consequently, this particular model of the distant solar system is rejectable at the 2--3$\sigma$ level. For the more massive and eccentric planet, we find that far too many highly inclined scattering objects are generated. The number of $i>45^{\circ}$ scatterers is $\sim$4--6 times too large, and the median inclination of all detectable scatterers is $\sim$twice as large as the observed value of 13.7$^{\circ}$. These results are consistent across four different assumed absolute magnitude distributions. 

Another notable feature of both distant planet models is that they generate a rather modest asymmetry in the directions of detached TNOs' longitude of perihelion.  Only $\sim$60--75\% of decoupled objects have $\varpi$ within $\pm90^{\circ}$ of the distant planet. Such a $\varpi$ distribution does not explain the observed $\varpi$ distribution among observed distant TNOs markedly better than an isotropic one. Moreover, the direction of the $\varpi$ asymmetry is very sensitive to TNO perihelion. In our models, there is an overabundance of perihelion longitudes that are anti-aligned with the distant planet for TNO perihelia between $\sim$35 and $\sim$50 AU. However, for TNO perihelia beyond $\sim$50 AU, our models predict the $\varpi$ asymmetry switches to favor orbits whose longitudes of perihelion are aligned with the planet, contrary to previous modeling efforts. (The magnitude of the $\varpi$ asymmetry is approximately the same in each perihelion regime of our models.) This asymmetry is due to our models' reliance on the distant planet to both detach orbital perihelia from the known planets and sculpt their $\varpi$ distribution. If such an assumption is made, then Sedna and 2012 VP$_{113}$ should not be included as evidence for the same $\varpi$ clustering observed among TNOs with perihelia below $\sim$50 AU. 

Although our simulated distant TNOs exhibit relatively mild $\varpi$ asymmetries, our simulations do not model the dynamical process of implanting a distant planet onto our assumed orbit, and it is possible that such a process could also implant a large population of TNOs decoupled from the known planets. This initially decoupled population of small bodies could then yield a $\varpi$ asymmetry more extreme than those seen in our work here. This does not resolve the problem of overly inclined scattering objects, but we find that a lower orbital inclination ($i=5^{\circ}$) for the distant planet may yield a scattering population that less strongly conflicts with observed scattering objects. Such distant planet models will have to be developed and tested more robustly in future works. Nevertheless, our present work shows that the orbits of TNOs actively scattering off the giant planets provide important information on the distant solar system, sampling the innermost regions of the Oort cloud and constraining the potential masses and orbits of any undiscovered distant planets.

\section{Acknowledgements}

This work was performed with support from NASA Emerging Worlds Grant 80NSSC18K0600, NSF award AST-1615975, a University of Oklahoma Junior Faculty Fellowship, and the University of Oklahoma Physics REU program (NSF Award 1359417). S.M.L. gratefully acknowledges support from the NRC-Canada Plaskett Fellowship. M.T.B. acknowledges support from UK Science and Technology Facilities Council grant ST/P0003094/1. The computing for this project was performed at the OU Supercomputing Center for Education \& Research (OSCER) at the University of Oklahoma (OU). We thank the anonymous reviewer whose comments and suggestions greatly improved the quality of this paper.

\bibliographystyle{apj}
\bibliography{HighIncCentaurs}

\end{document}